\documentclass{emulateapj}

\def\msol{\hbox{$\rm\thinspace M_{\odot}$}}  \def\etal{{\it et al\ }}
\def\eg{{\it e.g.\ }}   
   
   \def\p3m{P${}^3$M}
\def\ap3m{AP${}^3$M} \def\-{{\em{---}}}

\newcommand{\be}{\begin{equation}}  \newcommand{\ba}{\begin{eqnarray}}
\newcommand{\ee}{\end{equation}}  \newcommand{\ea}{\end{eqnarray}}

 \newcommand{\bi}{\begin{itemize}}
\newcommand{\ei}{\end{itemize}}

\def\lesssim{\mathrel{\hbox{\rlap{\hbox{\lower4pt\hbox{$\sim$}}}\hbox{$<$}}}}
\def\gtrsim{\mathrel{\hbox{\rlap{\hbox{\lower4pt\hbox{$\sim$}}}\hbox{$>$}}}}

\usepackage{epstopdf} \usepackage{subfigure} \usepackage{rotating} \usepackage{amsmath}

\begin{document}

\title{Formation of Compact Stellar Clusters by High-Redshift Galaxy Outflows III:  Observability and Connection to Halo Globular Clusters}
\author{William J. Gray\altaffilmark{1} \& Evan Scannapieco\altaffilmark{1}}

\altaffiltext{1} {School of Earth and Space Exploration, Arizona State University, P.O. Box 871404, Tempe, AZ, 85287-1494.}

\begin{abstract}

The early universe hosted a large population of low-mass virialized ``minihalos," that were not massive enough to form stars on their own.  While most minihalos were  photoevaporated by ionizing photons from star-forming galaxies, these galaxies  also  drove large outflows, which in some cases would have reached the minihalos in advance of ionization fronts. In the previous papers in this series, we carried out high-resolution, three-dimensional adaptive mesh refinement simulations of outflow-minihalo interactions that included non-equilibrium chemistry, radiative cooling, and turbulent mixing. We found that, for a fiducial set of parameters,  minihalos were transformed into dense, chemically homogenous stellar clusters.  Here we conduct a suite of  simulations that follow these interactions over a wide range of parameters including minihalo mass, minihalo formation redshift, outflow energy, outflow redshift, distance, concentration, and spin.  In almost all cases, the shocked minihalos form molecules through nonequillibrium reactions and then cool rapidly to become compact, chemically-homogenous stellar clusters.  Furthermore, we show that the unique properties of these clusters make them a prime target for direct study with the next generation of telescopes, and that there are many reasons to suspect that their low-redshift counterparts are the observed population of halo globular clusters.

\end{abstract}  


\section{Introduction}

	In the modern  theory of cosmic evolution, structures form hierarchically through accretion and mergers (\eg White \& Rees 1978; White \& Frenk 1991; Kauffmann \etal 1993; Cole \etal 2000; Bower \etal 2006).   A generic feature of this history, then, is the presence of a high-redshift population of gravitationally-bound structures that were much less massive than present-day galaxies.  At virial temperatures below 10$^{4}$ K, atomic transitions of hydrogen and helium are not excited, and the gas must radiate energy via molecular transition lines or dust emission. Although some H$_2$ left over from recombination was able to cool the earliest structures (\eg Abel \etal 2002; Bromm \etal 2002; Stacy \etal 2010; Turk \etal 2009), the resulting 11.2-13.6 eV background emission from these objects (\eg Haiman \etal 1997,2000; Ciardi \etal 2000; Sokasian \etal 2004; O'Shea \& Norman 2007) quickly dissociated the already low abundance of these primordial molecules (Galli \& Palla 1998). In fact, even if some small fraction of molecules survived, it is unlikely  to have had a large effect on the structure of these low-mass objects (\eg Whalen \etal 2008a; Ahn \etal 2009). The result of this inefficient cooling then, was a large population of primordial, atomic ``minihalos" that were unable to form stars until triggered by some exterior influence.

	A prime candidate for inducing star formation in these objects is galaxy outflows, powered by core-collapse supernovae and winds from massive stars. Such outflows have been found originating from a variety of galaxies, from dwarfs to massive starbursts, over a wide range of redshifts (\eg  Lehnert \& Heckman 1996; Franx \etal 1997; Pettini \etal 1998; Martin 1999; Heckman \etal 2000; Veilleux \etal 2005; Rupke \etal 2005; Chung \etal 2011). Furthermore, theoretical studies have argued that many of these early galaxies represent the tail end of a large population of low-mass starbursts that occurred before reionization (Scannapieco, Ferrera, \& Madau 2002; Thacker, Scannapieco, \& Davis 2002). While ionizing photons generated by  these early galaxies would have lead to the photoionization of large regions of space,  these photons are also easily trapped behind outflows as they sweep up intergalactic gas (Fujita \etal 2003). This suggests that many regions in the intergalactic medium may have be impacted by outflows before they were ionized.
	
	In the first two papers in this series (Gray \& Scannapieco 2010, hereafter Paper I; Gray \& Scannapieco 2011, hereafter Paper II) we carried out high-resolution, three-dimensional adaptive mesh refinement simulations of the interaction between a typical primordial minihalo  and high redshift galaxy outflow, capturing all the important physical processes in detail. In Paper I we looked at the influence of a 14-species primordial non-equilibrium chemical network with associated cooling terms, and showed that much of the minihalo baryonic matter  is removed from the dark matter halo and compressed into several dense clumps embedded in a ribbon of gas. The inclusion of a dissociating UV background, while altering the final molecular abundances, had no effect on the final distribution. In Paper II we extended this study by including the effects of metal-line cooling and turbulence. Turbulence had the primary effect of mixing metals into the primordial cloud from the outflow, which enriched the cloud to Z $\approx$ 10$^{-2}$ Z$_{\odot}.$ While metals provided another source of cooling, we found that they did not have a significant impact on the final structure of the cloud, which was very similar to the that found in Paper I. In both studies, dense $\approx 10^{5}$ \msol \ clusters were formed.
	 
	Additionally, these clusters had many of the properties of present day halo globular clusters. In particular, the halo globular cluster mass function is well defined by a Gaussian with a mean mass of 10$^{5}$ M$_{\odot}$ and dispersion of 0.5 dex (\eg Armandroff 1989), which spans the range of clusters formed in our studies. Furthermore, while the stars in a typical globular cluster have a small scatter in their metallicities,  the metallicity distribution function of halo globular clusters is also well defined by a Gaussian with a mean value of $\left[ \frac{Fe}{H} \right]$ $\approx$ -1.6 and a dispersion of 0.3 dex (Zinn 1985; Ashman \& Bird 1993). The clusters we formed in Paper II were enriched homogeneously via turbulent processes to approximately this mean value. Finally, our models naturally reproduced observations that suggest that globular clusters do not reside within dark matter potentials (Moore 1996; Conroy \etal 2010), as the outflows always removed the halo gas from their associated dark matter.
			
	These many tantalizing clues hint that outflow-minihalo interactions may in fact have provided the unknown mechanism that formed the observed population of halo globular clusters.  However, a proper assessment of this connection is not possible without understanding how robust the mechanism is to the many likely variations in the properties of high-redshift minihalos and galaxy outflows.  Thus, here we turn our attention to the  range of  interactions that lead to the formation of compact stellar clusters, carrying out a detailed parameter study in which we vary a wide range of properties, including  minihalo mass, minihalo formation redshift, outflow energy, outflow redshift, distance, concentration, and spin. With these results we are able to asses in detail the viability of this mechanism in producing halo globular clusters, as well as make a series of high-redshift predictions to be verified with the next generation of telescopes. 
	
	The structure of the paper is as follows. In Section \S2 we describe the model setup and parameters, and \S3 we present the results of these simulations. Finally, in \S4 we present simulated images of narrow band Ly${\alpha}$ observations from the next generation of ground based telescopes and wide band imaging with the {\em James Webb Space Telescope} ({\em JWST}) and  relate our models to present-day halo globular clusters. Our conclusions are given in \S5.

\section{Model Framework}

\subsection{Numerical Methods}

	All simulations were performed with FLASH version 3.1.1, a multidimensional adaptive mesh refinement (AMR) hydrodynamics code (Fryxell \etal 2000) which solves the Riemann problem on a cartesian grid using a directionally-split Piecewise-Parabolic Method (PPM) solver (Colella \& Woodward 1984; Colella \& Glaz 1985; Fryxell, M\"uller, \& Arnett 1989). In addition to making use of the efficient multigrid self-gravity package included with FLASH (Ricker 2008), we have developed a variety of other physics packages necessary for simulating outflow-minihalo interactions.
	
	First, we have implemented a version of the non-equilibrium chemistry network from Glover \& Abel (2008). This tracks the full set of two-body reactions that determine the evolution of three atomic hydrogen species (H, H$^+$, and H$^-$), three atomic helium species (He, He$^{+}$, and He$^{++}$), three atomic deuterium species (D, D$^{+}$, and D$^{-}$), two states of molecular hydrogen (H$_2$ and H$_2^+$), two states of hydrogen deuteride (HD and HD$^{+}$), and electrons (e$^{-}$). However, we have neglected the contribution from three-body reactions, since they only become important at densities of $n >$ 10$^{8}$ cm$^{-3}$ (\eg Palla \etal 1983), and neglected the contribution from reactions concerning molecular deuterium (D$_2$), because the cooling from D$_2$ and D$_2^+$ is negligible (Glover \etal 2008).  While we included UV dissociation rates from  Glover \& Savin (2009)  to capture the effect of a dissociating background, we discovered that for all reasonable choices of background intensity, they did not effect the evolution of these interactions significantly (Paper I).  Thus we did not include a dissociating background in the present study.  Further details of this implementation and  the impact of the UV background are given in Paper I. 
	
	Secondly, we have implemented molecular and atomic cooling rates. At temperatures below 10$^{4}$ K cooling is dominated by both molecular line transitions from H$_2$ and HD and metal line transitions. We have implemented the cooling rates from Glover \& Abel (2008) for the collisional excitation between H$_2$ and a variety of atomic species as well as the Lipovka \etal (2005) rate between HD and H.   Above 10$^4$ K, primordial cooling is dominated by atomic hydrogen and helium, whose cooling rates are calculated using CLOUDY (Ferland \etal 1998) assuming Case B recombination.  The metal-line cooling function is taken from Wiersma \etal (2009) via a table lookup, assuming solar abundances ratios, scaled by the local metal abundance. 
			
Finally,  we have implemented a buoyancy and shear driven model of turbulence that extends the two equation $K-L$ model of Dimonte \& Tipton (2006; see also Chiravalle 2006), where $K$ represents the turbulent kinetic energy and $L$ is the eddy length scale. This model reproduces the effect of three primary instabilities: the Rayleigh-Taylor instability that arises when a low density fluid supports a high density fluid under an acceleration, the Richtmyer-Meshkov instability, which appears when a shock interacts with a fluid with a different impedance (such as a density gradient), and the Kelvin-Helmholtz instability, which occurs between two fluids that shear in a direction perpendicular to their interface. This model also includes diffusion terms for both the molecular species and metal abundances, which allows for metal mixing. Further details of this package are given in Paper II.
	
	As in  Paper I and Paper II, we assume a standard $\Lambda$CDM cosmology with  $h$ = 0.7, $\Omega_0$ = 0.3, $\Omega_{\Lambda}$ = 0.7, and $\Omega_b$ = 0.045 (\eg Spergel \etal 2007), where $h$ is the Hubble constant with units of 100 km s$^{-1}$ Mpc$^{-1}$,  $\Omega_0$, $\Omega_{\Lambda}$, $\Omega_b$ are the total matter, vacuum, and baryonic matter densities, in units of the critical density. With our choice of $h$, the critical density is $\rho_{\rm crit}$ = $9.2 \times 10^{-30}$ g cm$^{-3}$. 

\subsection{Minihalo}

	In all runs the minihalo was comprised of a dark matter halo and a metal-free, neutral atomic cloud made up of  76\% hydrogen and 24\% helium by mass. The total mass of the minihalo was defined as M$_{c}$ $=$ M$_6 \times 10^{6}$ M$_{\sun},$ and we assumed that the dark matter and gas have collapsed by a redshift $z_c$ at which time the object had a mean overdensity of $\Delta$ = 178 (e.g., Eke \etal 1998). This leads to a mean density of $\rho_c = $ $\Delta$ $\Omega_0$ $(1+z_c)^3$ $\rho_{\rm crit}$ in the cloud. With these choices the virial radius of the cloud is
\be
	R_c = 0.3 M_6^{1/3} \left(\frac{1+z_c}{10}\right)^{-1} \ {\rm kpc},
\label{erc}
\ee	
and its virial velocity is
\be
	v_c = 4.4 M_6^{1/3} \left( \frac{1+z_c}{10} \right)^{1/2} \ {\rm km \ s^{-1}}.
\label{evc}
\ee

The radial profile of the minihalo was  taken from Navarro \etal (1997):
\be
	\rho(R) = \frac{ \Omega_0 \rho_c}{cx(1+cx)^2} \frac{c^2}{3 F(c)} \ {\rm g \ cm^{-3}},
\label{enfw}
\ee
where $c$ is the halo concentration parameter, $x \equiv R/R_c$, and $F(t)$ $\equiv$ $\ln(1+t) - \frac{t}{1+t}$. As the minihalo collapses, we assume that the gas is isothermal and heated to its virial temperature of
$T_c = 720 M_6^{2/3} [(1+z_c)/10] \ {\rm K},$ 
which produces an isothermal  gas density distribution in the CDM potential well of
\be
	\rho_{\rm gas}(R) = \rho_0 e^{-[v_{\rm esc}^2(0) - v_{\rm esc}^2(R)]/v_c^2} \ {\rm g \ cm^{-3}},
\label{rhogas}
\ee  
where $v_{\rm esc}^2(R = xR_c) = 2 v_c^2 [F(cx) + cx/(1+cx)] [xF(c)]^{-1}$
is the escape velocity of a particle at a distance $R$ from the halo center, $v_{\rm esc}^2(0) = 2 v_c^2 c/ F(c)$, and $c$ is the halo concentration factor which has a fiducial value of 4.8 (Madau \etal 2001). 
The central gas density of the cloud is then 
\begin{eqnarray}
	\rho_0 &=& \frac{(178/3) c^3 \Omega_b e^A (1+z_c)^3}{\int_0^c(1+t)^{(A/t)} t^2 dt}\   {\rm gm\  cm^{-3}}, 
\end{eqnarray}
where $A \equiv 2c/F(c)$ and $t \equiv c x.$ 
Outside the virial radius of the cloud, we assume that the gas is at the same temperature as the minihalo and in hydrostatic balance with it. This leads to a density distribution of 
\be
	\rho(R > R_c) = \rho(R_c) e^{\frac{R_0}{R} - \frac{R_0}{R_c}} \ {\rm gm\  cm^{-3}},
\label{erz}
\ee
where $R$ is the radius, $R_c$ is the virial radius, and $R_0$ = $G M_c m_p / k_b T$, where M$_c$ and $T$ are the mass and virial temperature of the cloud.  

We begin each simulation with the halo in hydrostatic equilibrium using a two part gravity scheme to account for the self-gravity of the gas and the dark matter halo. The self-gravity was handled by the efficient multigrid Poisson solver (Ricker 2008) which is included in FLASH.  The dark matter term was handled by first calculating the total gravitational acceleration due to the total (dark matter + gas) mass  distribution (via Eqn.~\ref{rhogas}) and subtracting off the term for the initial gas configuration before adding the results of the self-gravity calculation. Thus, initially when the cloud is pressure supported against collapse the gas-only and self-gravity terms cancel.  Outside the acceleration due to the dark matter was calculated as $- G M_c / R^2$ cm s$^{-2},$ and again the gas was taken to be in initial hydrostatic equilibrium according to Eq.\ \ref{erz}.

\subsection{Outflow}

The galactic outflow was modeled as a Sedov-Taylor blast wave. We assumed that the minihalo sits at a (physical) distance $R_s$ and that the shock moves with a velocity of
\be
	v_s = 760 \delta_{44}^{-1/2} (\epsilon E_{55})^{1/2} \left( \frac{1+z_s}{10} \right)^{-3/2} R_s^{-3/2} \ {\rm km \ s^{-1}},
	\label{sevel}
\ee
where $\delta_{44} = \delta/44$, and $\delta$ is defined as the ratio of the density of the gas compared to the mean density at that redshift, $\epsilon E_{55}$ is the input energy of the shock wave where $\epsilon$ is the wind efficiency and $E_{55}$ is the total SNe energy in units of 10$^{55}$ ergs, and $z_s$ is the redshift when the outflow reaches the halo  (Scannapieco \etal 2004). The postshock temperature is 
$T_s = 1.4 \times 10^5 (v_s/100 \ {\rm km \ s^{-1}})^2 \ {\rm K}$. By the time the outflow reaches $R_s$ it has entrained a mass of
\be
	M_{\rm s,total} = 1.4 \times 10^6 \delta_{44} \left( \frac{1+z_s}{10} \right)^3 R_s^3 \ {\rm M_{\sun}},
\ee
and has a surface density of 
\be 
	\sigma_s = 1.0 \times 10^5 \delta_{44} \left( \frac{1+z_s}{10} \right)^3 R_s \ {\rm M_{\sun} \ kpc^{-2}}.
\ee

The outflow is assumed to consist of completely ionized hydrogen and helium with the same relative abundances as the primordial gas, but enriched with metals from the supernovae that drove the material out of the host galaxy. To determine the metal abundance, we followed the estimate from Paper II and Scannapieco \etal (2004) and assumed that each supernova, whether a core-collapse or more-exotic pair-instability supernova (Woosley \& Weaver 1995; Heger \& Woosley 2002), generates 2 M$_{\sun}$ of metals per $10^{51}$ ergs of energy. If half of these metals are funneled into the outflow, we can expect a total mass of metals of M$_{\rm metal} = 10^4 E_{55}$ M$_{\odot}$. We therefore initialized the shock with an initial abundance of 
\be
	\frac{Z_{\rm metal}}{Z_{\odot}} = \frac{M_{\rm metal} }{ M_{\rm s,total}} = \frac{10^4 E_{55} \ M_{\odot}}{1.4 \times 10^6 \delta_{44} \left( \frac{1+z_s}{10} \right)^3 R_s^3}
	\label{shockmetal}
\ee
The lifetime of the shock can be estimated from $\sigma_s$ = $\rho_{\rm post}$$v_{\rm post}$t$_{s}$, where $\sigma_s$ is the surface density of the outflow, $\rho_{\rm post}$ is the post shock density, $v_{post}$ is the post shock velocity, and t$_s$ is the shock lifetime. After a period t$_{full}$, which is nominally defined as 1.5 Myrs, the shock was allowed to taper off by slowly lowering the density and raising the temperature so that the pressure stayed constant. This prevents the courant time step from becoming exceedingly small long after the shock had passed over the minihalo. For more setup details see Paper I and Paper II. 

\begin{table*}
\begin{center}
\caption{Summary of Study Parameters. M$_6$ is the minihalo mass in units of M$_6$ = M$_c/10^6$ M$_{\odot}$, E$_{55}$ is the energy of the shock in units of E/$10^{55}$ ergs,  R$_s$ is the distance between the galaxy and the minihalo in units of (physical) kpc, z$_c$ is the redshift at which the halo virializes, z$_s$ is the redshift at which the shock reaches the minihalo,  $\lambda'$ is the spin parameter (see \S \ref{hs}), $c$ is the concentration parameter, Z is the metal abundance of the outflow in units of solar metallicity ({\it{Z$_{\odot}$}}), and Res is the minimum resolution of each simulation in units of pc. }
\begin{tabular}{ccccccccccccc}
\hline
      Name &  M$_6$ &  E$_{55}$ & R$_s$ & z$_c$ & z$_s$ &  $\lambda'$ &  c & Z &   Res & Notes \\
                  &                &                    & (kpc)      &             &             &                      &      & (Z$_\odot$)   & (pc)    & \\
\hline
      OFID	&	3 	&         10 &             3.6 &         10 &          8 &          0 &        4.8 &      0.12 &   9.14    & Original Fiducial \\
      PM10/NFID	&	10 	&  10 &           3.6 &         10 &          8 &          0 &        4.8 &      0.12 &   19.85  &     New Fiducial \\
      PM03	&	0.3 	&         10 &             3.6 &         10 &          8 &          0 &        4.8 &      0.12 &   5.06    &  \\
      PM30	&	30 	&         10 &             3.6 &         10 &          8 &          0 &        4.8 &      0.12 &   19.85  &  \\
      PE1    &	10 	&         1  &               3.6 &         10 &          8 &          0 &        4.8 &      0.06 & 19.85  & \\
      PE5	&	10 	&         5  &               3.6 &         10 &          8 &          0 &        4.8 &      0.06 & 19.85  & \\
      PE20	&	10	&         20 &             3.6 &         10 &          8 &          0 &        4.8 &      0.24 & 19.85  &  \\
      PE30	&	10	&         30 &             3.6 &         10 &          8 &          0 &        4.8 &      0.36 &   19.85  &  \\
      PZC8	&	10 	&         10 &             3.6 &           8 &          8 &          0 &        4.8 &      0.12 &   19.85  & \\
      PZC15 & 	10	&         10 &             3.6 &         15 &          8 &          0 &        4.8 &      0.12 &   19.85  & \\
      PZS10 &      10     &	  10 &	     3.6&	     10 &	     10 &	     0 &       4.8 &      0.12  &  19.85  & \\
      PR21	&	10 	&         10 &             2.1 &         10 &          8 &          0 &        4.8 &      0.60 &  19.85  & \\
      PR66	&	10 	&         10 &             6.6 &         10 &          8 &          0 &        4.8 &      0.019 &  19.85  & \\
      PR120	&  	10 	&         10 &              12 &         10 &          8 &          0 &        4.8 &      0.003 &  19.85  &  \\
      PSPZ  &  	10 	&         10 &             3.6 &         10 &          8 &     0.023 &        4.8 &      0.12  & 19.85  & Angle = 0$^\circ$ \\
      PSPN &  	10	&         10 &             3.6 &         10 &          8 &     0.023 &        4.8 &      0.12  & 19.85  & Angle = 90$^\circ$  \\
      PC32	&	 10	&         10 &             3.6 &         10 &          8 &          0 &         3.2 &      0.120 & 19.85  &  \\
      PC73	&  	10 	&         10 &             3.6 &         10 &          8 &          0 &         7.3 &      0.12  & 19.85  &  \\
\label{params}
\end{tabular} 
\end{center}
\end{table*} 

\section{Parameter Study}
\label{sps}
Table~\ref{params} summarizes the range of parameters studied in our simulations. Each case is given a name based on the value of the parameter that was changed from the fiducial case. OFID is the fiducial model from Paper II, in which it was labeled MRWT. In all simulations the wind efficiency was $\epsilon=0.3,$  and the mean overdensity of the medium between the outflow and the minihalo was 44 ($\delta_{44}=1$).  Each simulation was performed in a rectangular box with the $x$-axis twice the size of the $y$- and $z$-axes. The minihalo was centered at [0,0,0] and the shock originated from the left $x$ boundary. The base grid was taken to have 16 by 8 by 8 cells in the $x,$ $y,$ and $z$ directions, and l$_{\rm max}=5$, such that up to 4 additional levels of refinement were  added in regions with significant pressure or density structure.  In Paper I and Paper II we varied the maximum resolution in our simulations, and showed that such medium-resolution ($256 \times 128 \times 128$ effective) simulations were able to faithfully reproduce the outcome of shock-outflow interactions.  In the interest of speed we also included a forced-derefinement routine that required regions to derefine if their density was less than 3.0$\times 10^{-26}$ g cm$^{-3}$. This was implemented outside 393 pc from the center of the halo and after 7 Myrs from the beginning of the simulations, such that it only impacted low-density regions far from the evolving cloud. 
 
Each simulation was run to a time at which the shock completely overran the minihalo, which was typically on the order of a few Myrs. However, the clouds are expected to evolve over hundreds of Myrs before reaching a final configuration, a time that is much longer than we were able to run our FLASH simulations. To get around this limitation, we transformed the final mass distribution in our 3-D  simulations into a 1-D ballistic problem, as described in Paper I. 

Here, the mass distribution at the final output from each simulation was divided into 100 evenly spaced bins along the $x$-axis, and the mass of each bin was calculated by summing together the gas within a cylinder with a radius of  $10^{21}$ cm and a length set by the bin spacing.  Next each bin was converted into a particle, whose initial position was located at the center of mass of the summed density distribution, and whose initial velocity was calculated from conservation of momentum. These particles were then  evolved ballistically using a leap-frog method. At every time step, the acceleration of each particle was calculated from the self-gravity from all other particles as well as from the gravity from the dark matter halo. If a particle moved past the one in front of it, they  were merged  by adding together their masses and calculating a new velocity from momentum conservation.  The outcomes from the different runs were then compared with each other at a time of 200 Myrs after the end of each simulation.

\begin{figure*}
\begin{center}
\includegraphics[scale=0.20]{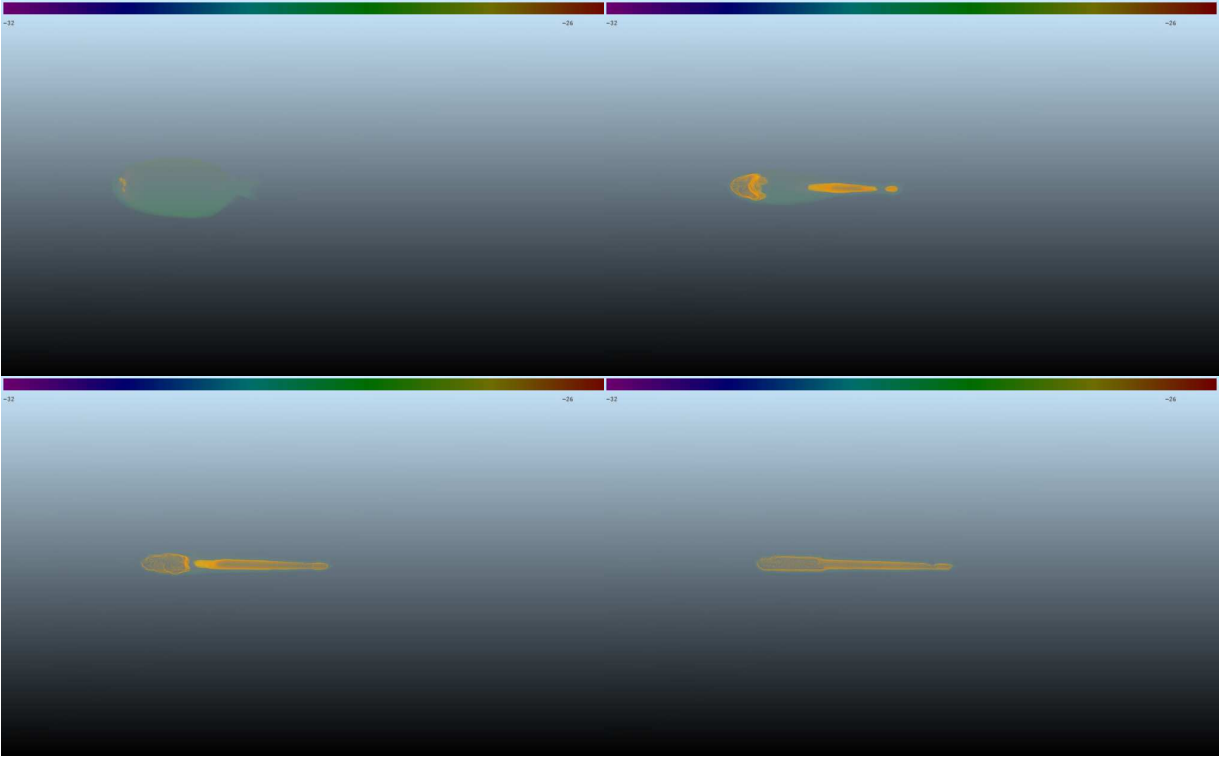}
\caption{Rendered snapshots of the logarithm of the H$_2$ density for run PM10/NFID.  Colors show logarithmic contours of H$_2$ density from 10$^{-32}$ to 10$^{-26}$ g cm$^{-3}$. {\em Top Left:} $t$ = 12.2  Myrs,  showing  the outflow  completely surrounding the minihalo and H$_2$ formation starting at the front of the cloud. {\em Top Right:} $t$ = 15.7 Myrs, showing the minihalo as it is collapsing and during which H$_2$ is increasing towards the center of the cloud. {\em Bottom Left:} $t$ = 19.2 Myrs, showing both the cloud as it is stretched and the H$_2$ that is then found throughout the dense ribbon. {\em Bottom Right:} $t$ = 22.0 Myrs, showing the final state of run PM10 and the uniformly cold ribbon of gas.}
\label{helios}
\end{center}
\end{figure*}

\subsection{Effect of Halo Mass}

	The outflow-minihalo interaction goes through several evolutionary stages, which are illustrated in Figure \ref{helios}. As the outflow impacts the front of the minihalo, the atomic gas is ionized but then cools rapidly.  As the temperature approaches $\approx 10^4$K,  molecules begin to form, catalyzed by H$^-$ and H$^+$. Since the shock moves faster through the less dense regions of the minihalo, molecule formation predominately occurs in the gas surrounding the center of the impacted minihalo (top left panel).  When the outflow meets at the back of the minihalo, however, a second shock is driven backwards toward the center of the collapsing cloud. This promotes molecule formation in the center of the minihalo, which continues to collapse as it further loses pressure support (top right panel).   As the cloud becomes more elongated (bottom left panel), the shear layer between the outflow and cloud becomes more and more turbulent, mixing metals into the cloud.  The final state of the simulation (bottom right panel) is similar to that seen in the runs in Paper I and Paper II: the gas from the minihalo has been moved out from the dark matter halo and formed a dense, cold ribbon of material stretched along the $x$-axis. 

\begin{figure*}
\begin{center}
\includegraphics[scale=0.50, clip, trim=50 0 0 0]{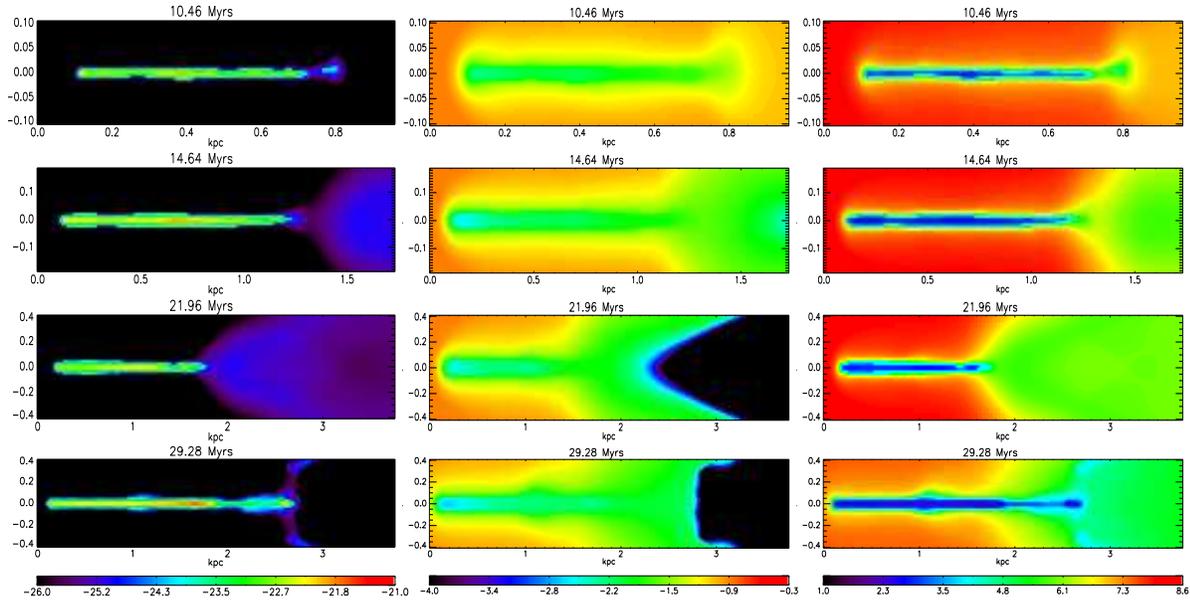}
\caption{Comparison of final outputs from runs with varying halo masses. The first column shows logarithmic density contours from 10$^{-26}$ to 10$^{-21}$ g cm$^{-3}$, the second column shows logarithmic metallicity contours between 10$^{-4}$ and 10$^{-0.3}$ Z$_{\odot}$, and the third column shows logarithmic temperature contours between 10 and 10$^{8.6}$ K.  The first row shows results from run PM03, the second row shows results from OFID, the third row shows results from run PM10, and the last row shows results from run PM30. In all cases the outcome is similar. The minihalo gas is moved out from the dark matter halo, stretched along the $x$-axis, and enriched homogeneously.}
\label{mmass}
\end{center}
\end{figure*}

	As the total mass in the impinging wind is relatively small,  the minihalo is the primary source of gas for the final dense stellar clusters.  Figure \ref{mmass} shows the impact of changing this mass, illustrating the final state of runs with minihalo masses ranging between 0.3 to 30 $\times 10^{6}$ M$_{\odot}$.  In all cases the gas from the minihalo forms dense cold clumps embedded in a ribbon of diffuse gas, stretching away from the dark matter halo.  The larger the halo mass, the longer the ribbon.  In the larger halos, a very slight metallicity gradient is observed as some of the gas in the center of the halo is not as enriched as the surrounding gas. However, every part of the ribbon is enriched to well above 10$^{-3} Z_{\odot},$ and the majority of the gas is enriched to a nearly constant value of  $\approx10^{-2} Z_{\odot}$. 
	
\begin{figure}
\includegraphics[scale=0.45]{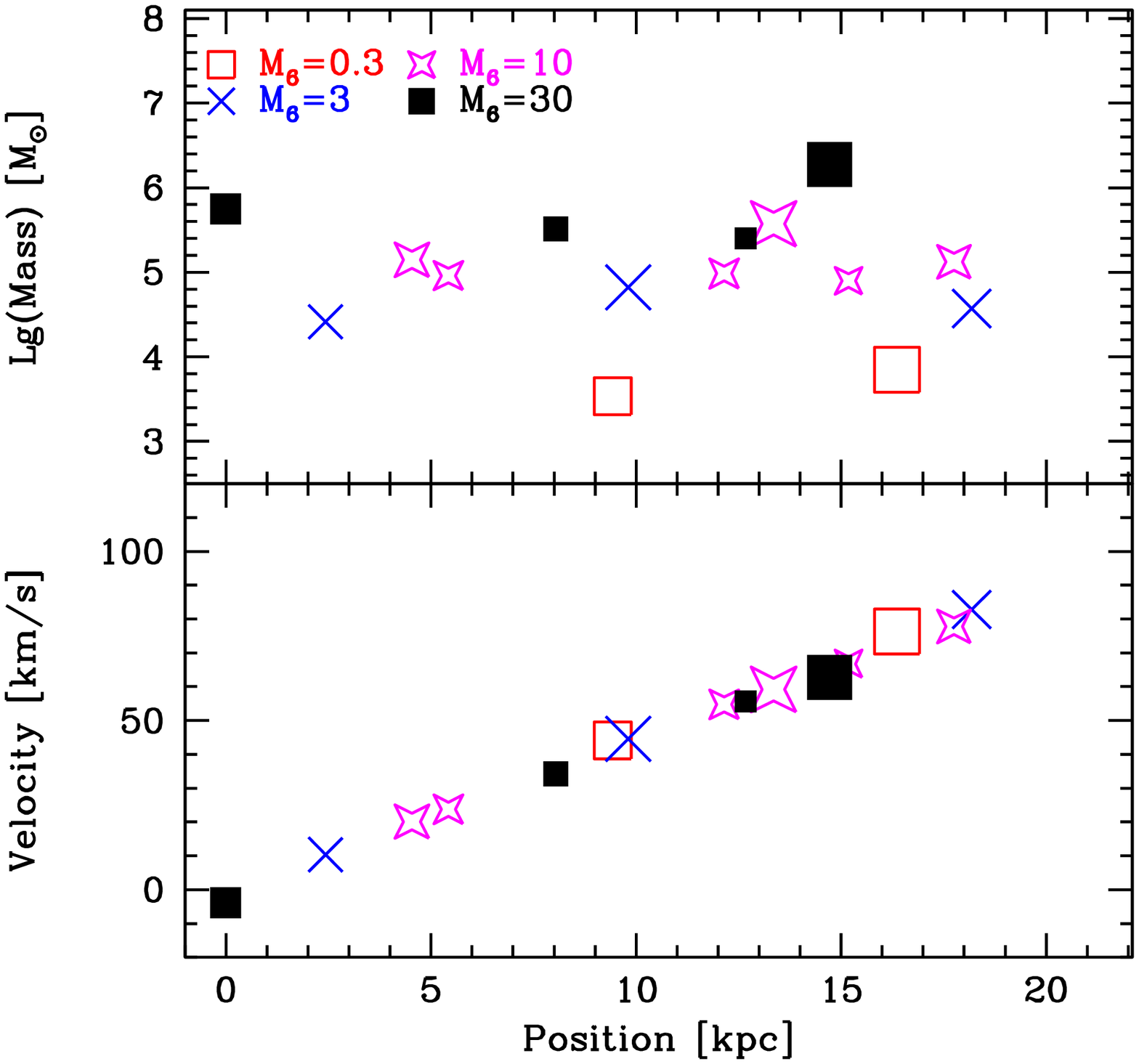}
\caption{Comparison of clusters generated in models with varying halo mass at a time of 200 Myrs after the end of the simulations. The top panel shows the logarithmic mass and the bottom panel shows the velocity of each cluster. The $x$-axis is the distance of each cluster from the center of their dark matter halos. The (red) unfilled squares show the PM03 model, the (blue) crosses show the OFID model, the (magenta) stars show the NFID/PM10  model, and the (black) filled squares show the PM30 model. The relative size of the points is proportional to the fraction of the total minihalo gas mass contained in each cluster. Only clusters that have  masses greater than 5\% of the initial baryonic halo mass are shown. The total mass of clusters from each model is 4.1 $\times 10^{4}$, 4.0 $\times 10^{5}$, 1.6 $\times 10^{6}$,  and 4.8 $\times 10^{6}$ \msol, for PM03, OFID, PM10, and PM30 respectively. Except for the high mass-case, every cluster is found outside of its dark matter halo.   }
\label{mass}
\end{figure}

	Figure \ref{mass} shows the evolved distribution of stellar clusters. The relative sizes of each cluster corresponds to their final mass, with larger symbols representing larger clusters. As expected, the more massive the initial halo, the more massive the final clusters. In the case of the largest mass minihalo, the outflow is not strong enough to remove all the gas.  Instead it leaves behind a rather sizable cluster at the center of the dark matter halo, although even more massive clusters are formed behind it.  In all other cases, the outflow effectively removes all of the gas and forms several dense clusters. Since the final cluster sizes are larger in the PM10 model relative to the $M_6=3$ run, but $10^7 M_\odot$ solar mass minihalos are nevertheless quite common at high-redshift, we have used it as our new fiducial model (NFID) and maintained this value in the simulations discussed below.

\subsection{Effect of Shock Energy}

	The shock plays three important roles in the evolution of these interactions. First, it accelerates the minihalo gas with enough momentum to move it out of the dark matter halo, second, it provides enough energy to ionize the gas, trigging the non-equilibrium chemical reactions that provide an avenue for H$_2$ and HD formation and cooling, and finally, the shock brings in the metals that enrich the post-shock star-forming medium. The single most important parameter in determining each of these effects is the energy  driving the shock, $E_{55}$, as this not only sets the shock velocity (Eq.~\ref{sevel}), but determines the total associated metal mass and abundance (Eq.~\ref{shockmetal}).  
	
	The fiducial estimate for $E_{55}$ is taken from Scannapieco \etal (2004) where they estimate the energy in the outflow of a high-redshift starburst galaxy. First, consider a a young galaxy of total mass $M_g$ = 10$^{9}$ M$_{\odot}$ that has a virial temperature greater than 10$^4$ K and is therefore allowed to cool via atomic hydrogen. If 10\% of the gas is converted into stars and 10$^{51}$ ergs of energy is produced by SNe for every 30 M$_{\odot}$ of stellar material for massive stars (Pop III, Heger \& Woosley 2002) then the expected total energy given off is $E_{55} \approx 50 M_{9}$, where $M_{9}$ is the mass of the galaxy in units of 10$^{9}$ M$_{\odot}$. In simulations by Mori \etal (2002) the wind efficiency $\epsilon$ is found to be $\approx 0.3$ for a $2.0\times10^{8} M_{\odot}$ starbursting galaxy. Using these numbers, the fiducial value for  $E_{55}$ is obtained. To get an idea of how this parameter alters the evolution of the minihalo, we allow this value to range between $1 < E_{55} < 30$. 

\begin{figure*}
\centerline{\includegraphics[scale=0.55,clip, trim=50 0 0 0 ]{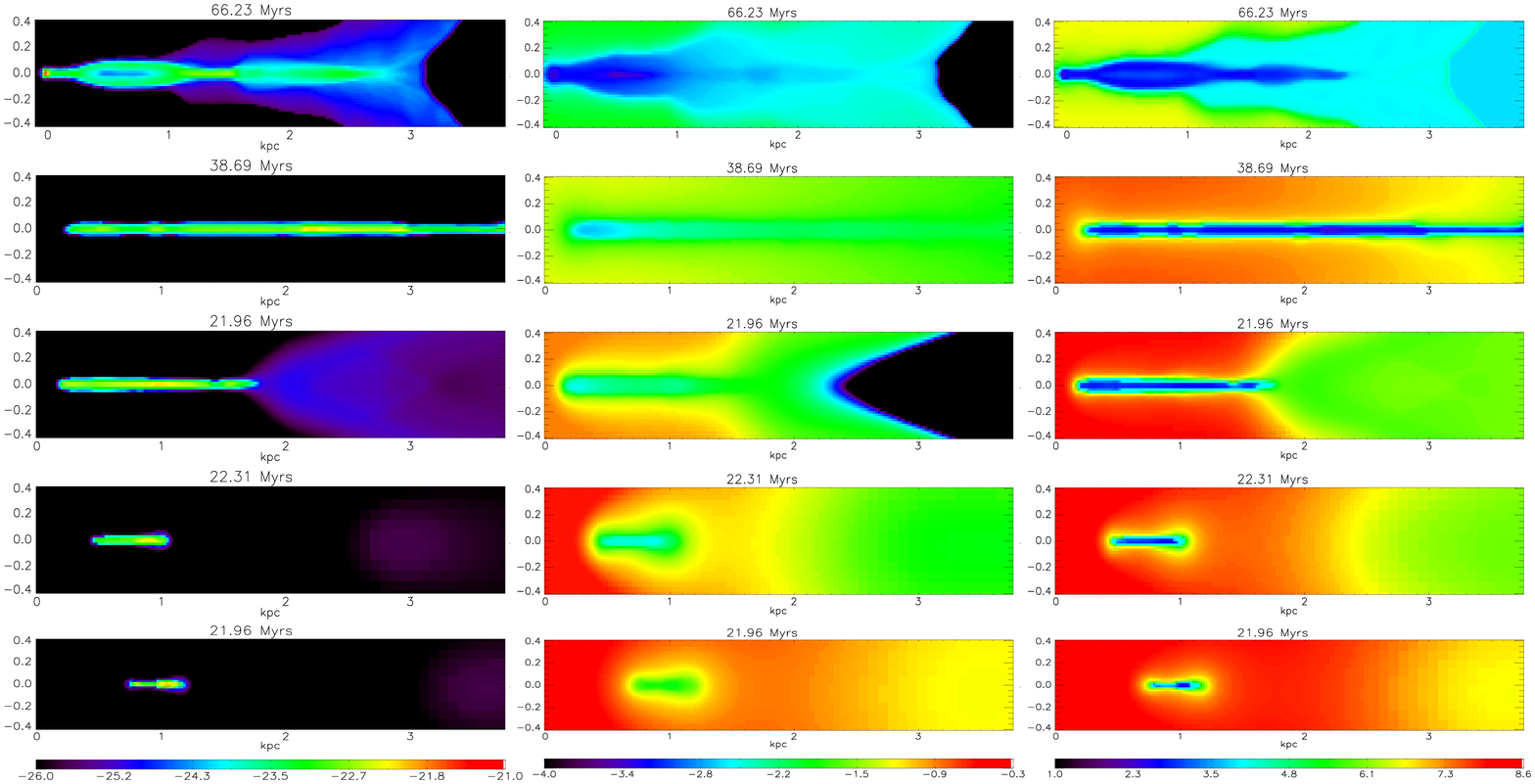}}
\caption{Comparison of final outputs from runs with varying outflow energies.  From top to bottom each row represents runs with $E_{55}$ = 1, 5, 10, 20, and 30, respectively.
As in Figure \ref{mmass} the first column shows logarithmic density contours from 10$^{-26}$ to 10$^{-21}$ g cm$^{-3}$ and the second column shows logarithmic contours of metallicity  from 10$^{-4}$ to 10$^{-0.3}$ Z$_{\odot},$ but now the third column shows logarithmic temperature contours between 10 and 10$^{8.6}$ K. The length of the ribbon is correlated with the initial shock energy: the smaller the energy, the longer the ribbon.}
\label{menergy}
\end{figure*}

	Figure \ref{menergy} shows the results of altering this key parameter. Each row corresponds to a different value of  $E_{55},$ which increases from top to bottom as 1, 5, 10, 20,  and 30, respectively.  As the shock energy increases,  the minihalo evolution changes dramatically.  Instead of stretching the gas into a ribbon as in the fiducial case, $E_{55}=10$ case, the stronger shocks crush the cloud into a single small, dense cluster. Interestingly, even though these high-energy interactions happen quickly, there is still enough time to mix in the metals from the shock.  In all cases the metallicity is $\approx 10^{-2} Z_{\odot},$  a value that is roughly constant across models  in part because the higher energy shock models have higher initial metal abundances.   On the other hand,  the lower the shock energy, the more the cloud is stretched into a diffuse ribbon in which  smaller clusters are embedded.  
	
\begin{figure}
\includegraphics[scale=0.45]{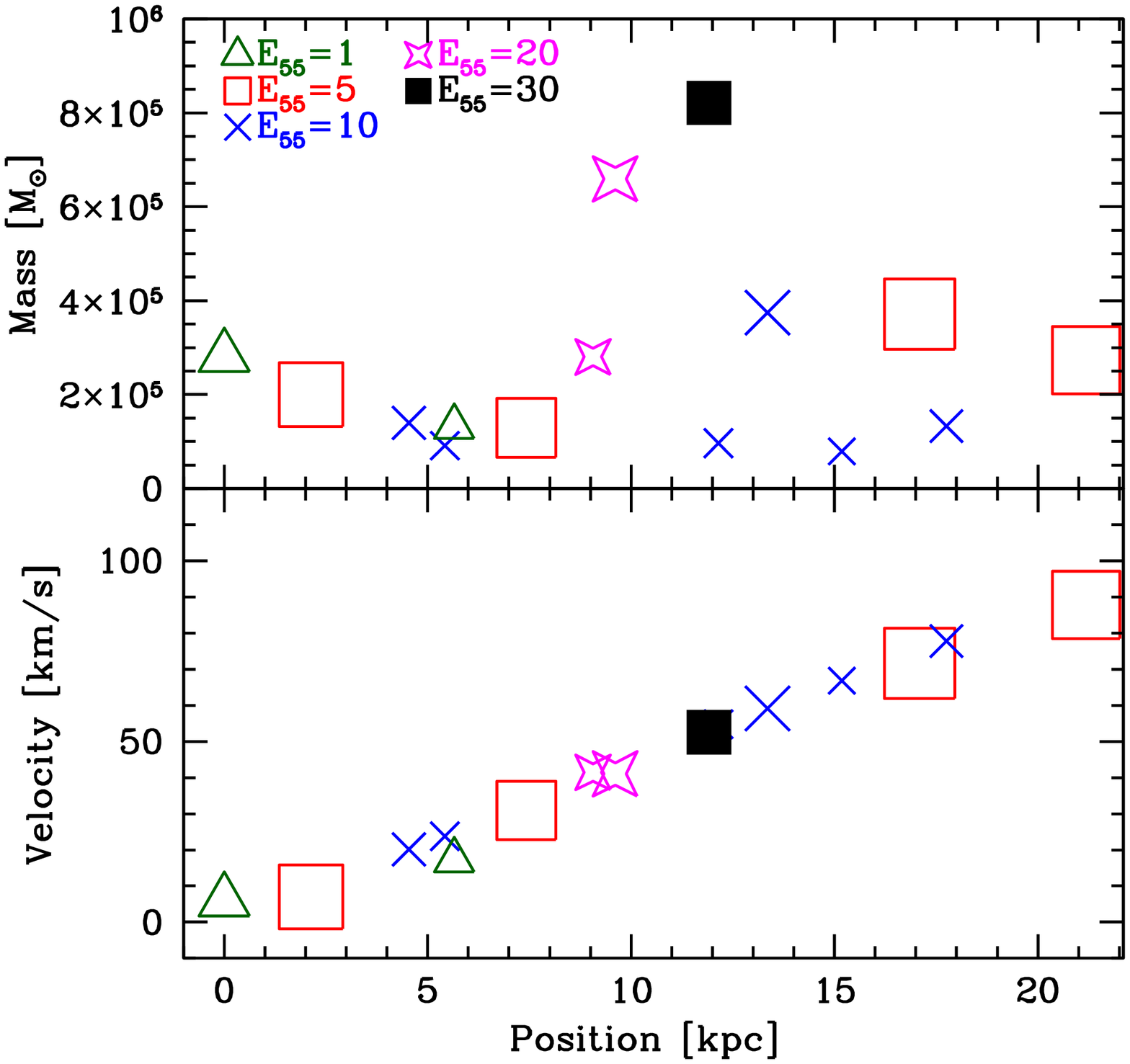}
\caption{Comparison of clusters generated in runs with  varying outflow energies. Panels are the same as Figure \ref{mass}, except the $y$-axis of the top panel is given in linear units rather than logarithmic. The (green) unfilled triangles show the clusters formed from PE1, (red) unfilled squares show the clusters from PE5, the (blue) crosses show the clusters from NFID, the (magenta) stars show the clusters from PE20, and the (black) filled squares show the clusters from PE30. The total mass of the clusters formed is 9.0 $\times 10^{6}$,  2.3 $\times 10^{6}$, 1.6 $\times 10^{6}$, 9.5 $\times 10^{5}$, and 8.2 $\times 10^{5} M_\odot$ in PE1, PE5, NFID, PE20, and PE30 respectively. In general, the larger the outflow energy, the more the initial halo is crushed rather than forming a long stream of gas. This leads to the formation of fewer, larger clusters.  }
\label{energy}
\end{figure}

	Over time, much of the surrounding gas merges with the initial cluster, which is similar to the evolution we found in Paper I and Paper II. However, this is not true in models with high-energy shocks since no extended ribbon is formed. These differences are apparent in Figure \ref{energy} which shows the final state of each model. Here we see that the mass in the high energy models is concentrated into a single object, moving outwards at a significant velocity from the dark matter halo.  Conversely, for low $E_{55}$ values, the gas condenses to form a number of smaller star clusters that are well separated from each other.  At the lowest energy level of $E_{55} = 1,$ the shock strips some of the baryonic matter from the dark matter halo but leaves behind a small dense cluster at the center of the halo. Additionally, the metal content of such a weak shock is not sufficient to enrich the resulting cluster to the $\approx 10^{-2} Z_{\odot}$ threshold.

	 
\subsection{Effect of Minihalo Virialization Redshift}

\begin{figure*}
\centerline{\includegraphics[scale=0.55,clip, trim=50 0 0 0 ]{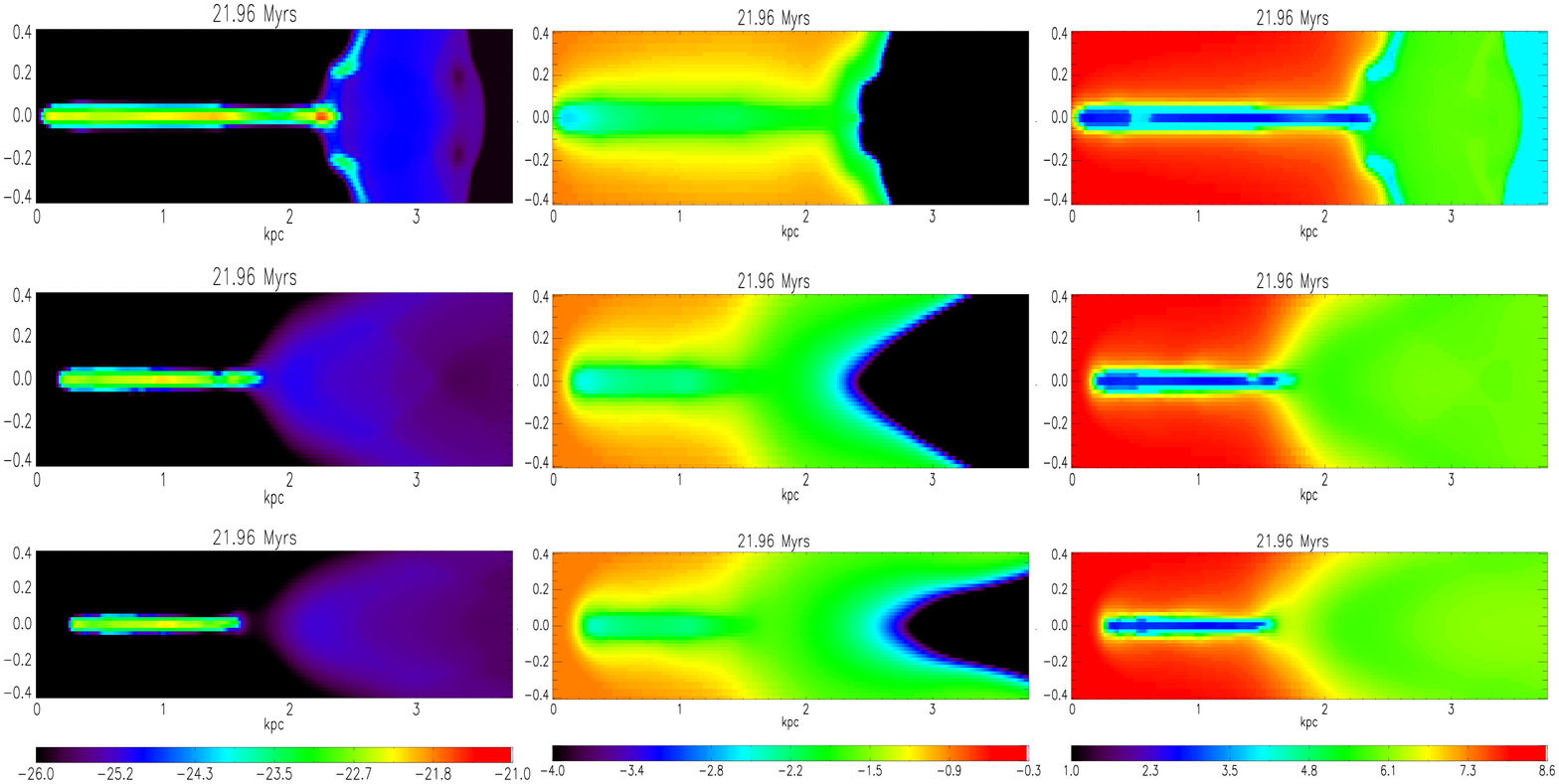}}
\caption{Comparison of  final outputs from runs with different minihalo virialization redshifts. 
 From left to right the columns show logarithmic contours of density, metallicity, and temperature with limits as in Figure \ref{menergy}. The top row shows the final state of run PZC15 ($z_c$ = 15), the middle row shows the final state of run NFID ($z_c$ = 10), and the bottom row shows the final state of run PZC8 ($z_c$ = 8). In highest $z_c$ redshift run, multiple large clusters are formed while in the lower $z_c$ runs, only one primary cluster is formed. In all runs, the metal abundance remains roughly constant and near Z $\approx$ 10$^{-2}$ Z$_{\odot}$ in the dense portions of the cloud, and the gas at the center of the dark matter halo is slightly deficient in metals.}
\label{zcmerge}
\end{figure*}

	Many of the basic properties of the halo, from the central density of the cloud to the radial density distribution, are strongly dependent on the minihalo virialization redshift (see Eqs.\ \ref{erc}-\ref{erz}). In general, the higher the virialization redshift, the more compact the minihalo, and thus the more resistant it is likely to be to external shocks.  To explore the impact of this parameter, we conducted two simulations in which $z_c$ was taken to be 10 (PZC10) and 15 (PZC15) as compared to our fiducial value of 8. The final states of each of these runs is shown  in  Figure \ref{zcmerge}.  It is important to note that the higher redshift simulations act as a more robust prediction of our model for globular cluster formation since by $z\sim8$ many of these minihalo may have been stripped of their baryons via ionization fronts during reionization. 
	
	For all virialization redshifts,  most of the mass from the halo is moved into a ribbon of material along the $x$-axis, but at higher virialization redshifts, the cloud is more compact (Eq.~\ref{erc}) and the gas in the center of the halo cannot escape as easily as the surrounding gas.   Also, as the virialization redshift increases, the physical size of the ribbon increases as does the total the entrained mass. Again, in all cases the clouds have been enriched to approximately 10$^{-2}$ Z$_{\odot}$ although the centers of the two oldest minihalos (those in runs PZC10 and PZC15) have slightly lower abundances. 

\begin{figure}
\includegraphics[scale=0.45]{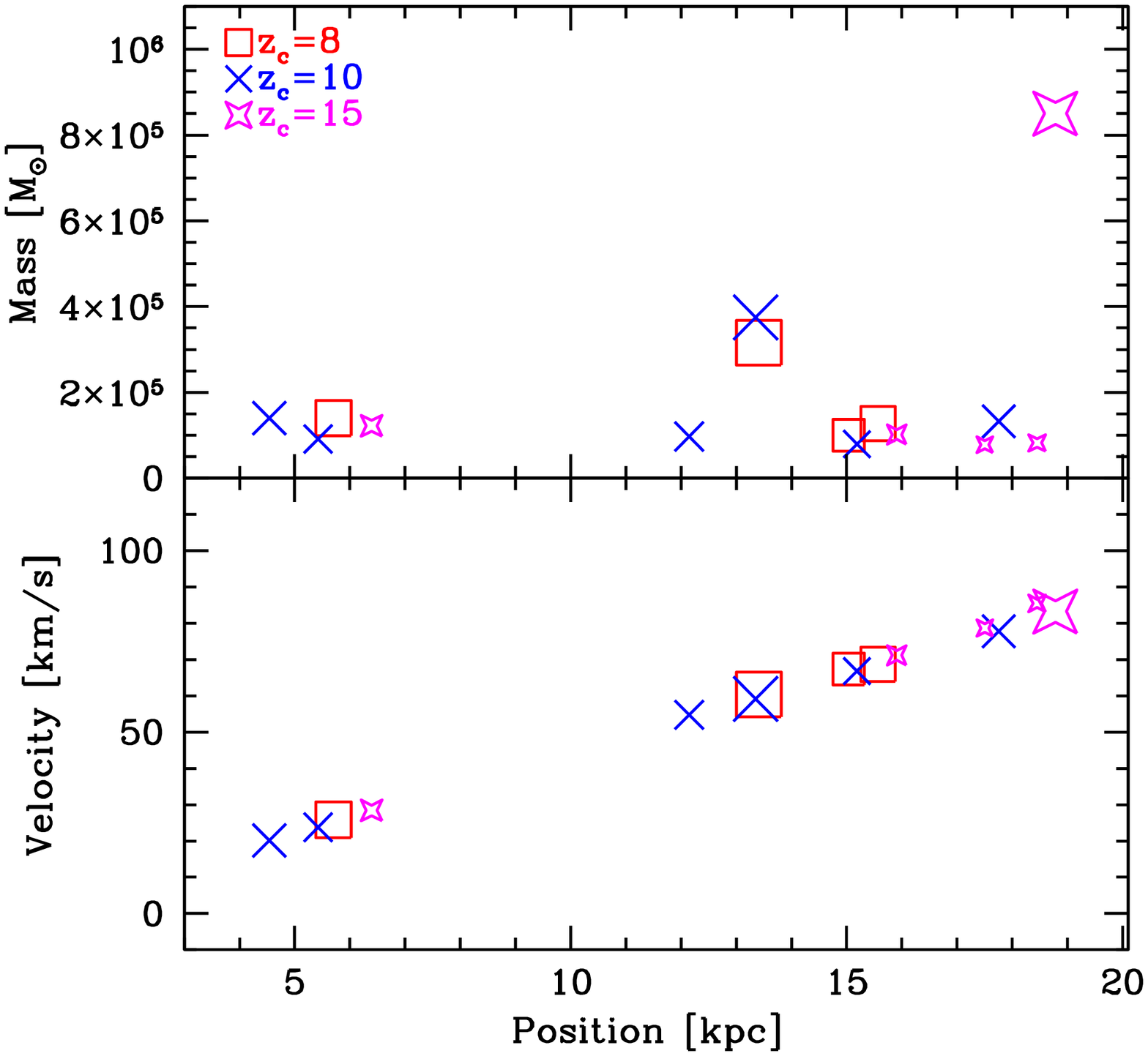}
\caption{Comparison of clusters generated in runs with different  virialization redshifts. Panels are the same as Figure \ref{energy}. The (red) unfilled squares show the clusters from PZC8, the (blue) crosses show the clusters from NFID, and the (magenta) stars show the clusters from PZC15. The total mass of clusters is 1.2 $\times 10^{6}$, 1.6 $\times 10^{6}$, and 2.3 $\times 10^{6}$ \msol\  for runs PZC8, NFID, and PZC15 respectively.  At lower redshifts the sizes and distribution of clusters is fairly similar, with at least one cluster with mass greater than 2.0 $\times 10^{5} M_\odot$ while at high redshift only one large cluster is formed. In all cases the largest clusters are found far from their respective dark matter halos. }
\label{zc}
\end{figure}

	Figure \ref{zc} shows the evolved distribution of these models. At lower virialization redshifts (z$_c$ = 8 and 10) the final distributions are very similar, and the number of clusters formed and the final mass of these clusters match well. At the highest virialization redshift, on the other hand, the distribution is quite different as one very large cluster is found nearly 20 kpc from the dark matter halo center. All of the other high redshift clusters are much smaller than those formed from minihalos with later virialization redshifts. This suggests that the formation of a large cluster (M $\approx$ few $\times 10^{5}$ \msol ) is robust over a large redshift range and may actually occur more easily for minihalos with earlier virialization redshifts. 

\subsection{Effect of Shock Redshift}

\begin{figure*}
\centerline{\includegraphics[scale=0.6,clip, trim=50 0 0 0 ]{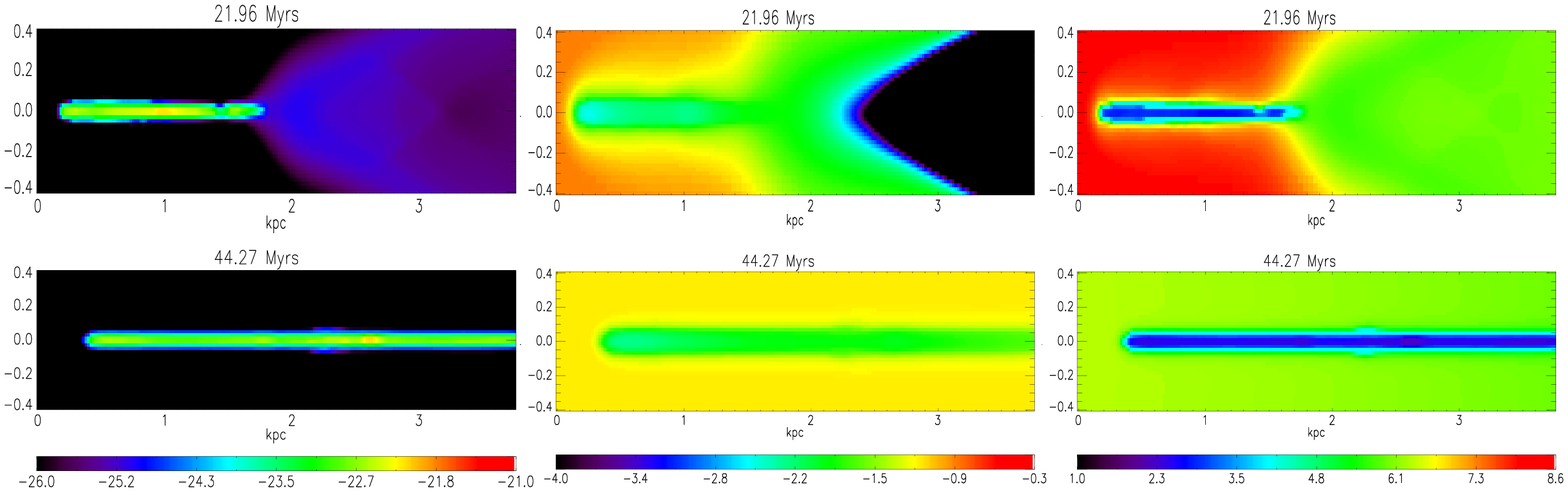}}
\caption{Comparison of final outputs from runs NFID (top) and PZS10 (bottom) which have outflow redshifts of $z_s=8$ and $z_s=10$, respectively. Panels are the same as Figure \ref{menergy}. The density is much more stretched and uniform than the fiducial case. While the metals from the shock have been mixed into the primordial gas, it is significantly more enriched than in the fiducial model. Both models reach the same final temperature.}
\label{zsmerge}
\end{figure*}

Other than $E_{55},$ the most important parameter in determining the properties of the shock is  its redshift, $z_s$. As the shock redshift is increased the mass in the shock increases (Eq.~\ref{sevel}) and the metal abundance decreases (Eq.~\ref{shockmetal}), which in turn affect all aspects of the interaction. To study the effect of a shock occurring at a higher redshift than our fiducial run, a model was run in which the outflow reached the minihalo at precisely the  minihalo virialization redshift (z$_s$ = z$_c$ = 10).  The final state of this run is contrasted with the fiducial case in Figure \ref{zsmerge}. 
	
	The increased mass in the shock in run PZS10 leads to a much more stretched, elongated, and uniform mass distribution than in the NFID run. This elongated distribution makes it easier for the metals carried by the shock to be efficiently mixed into the collapsing cloud, and thus the metal abundance of the cloud in this run is higher than the fiducial model, even though the abundance of the shock is lower.  On the other hand, the stretching in run PZS10 is not so severe as to suppress the formation of clusters, such as the one clearly visible at a distance of $\approx$2.5 kpc from the center of the dark matter halo.	
	
\begin{figure}
\includegraphics[scale=0.45]{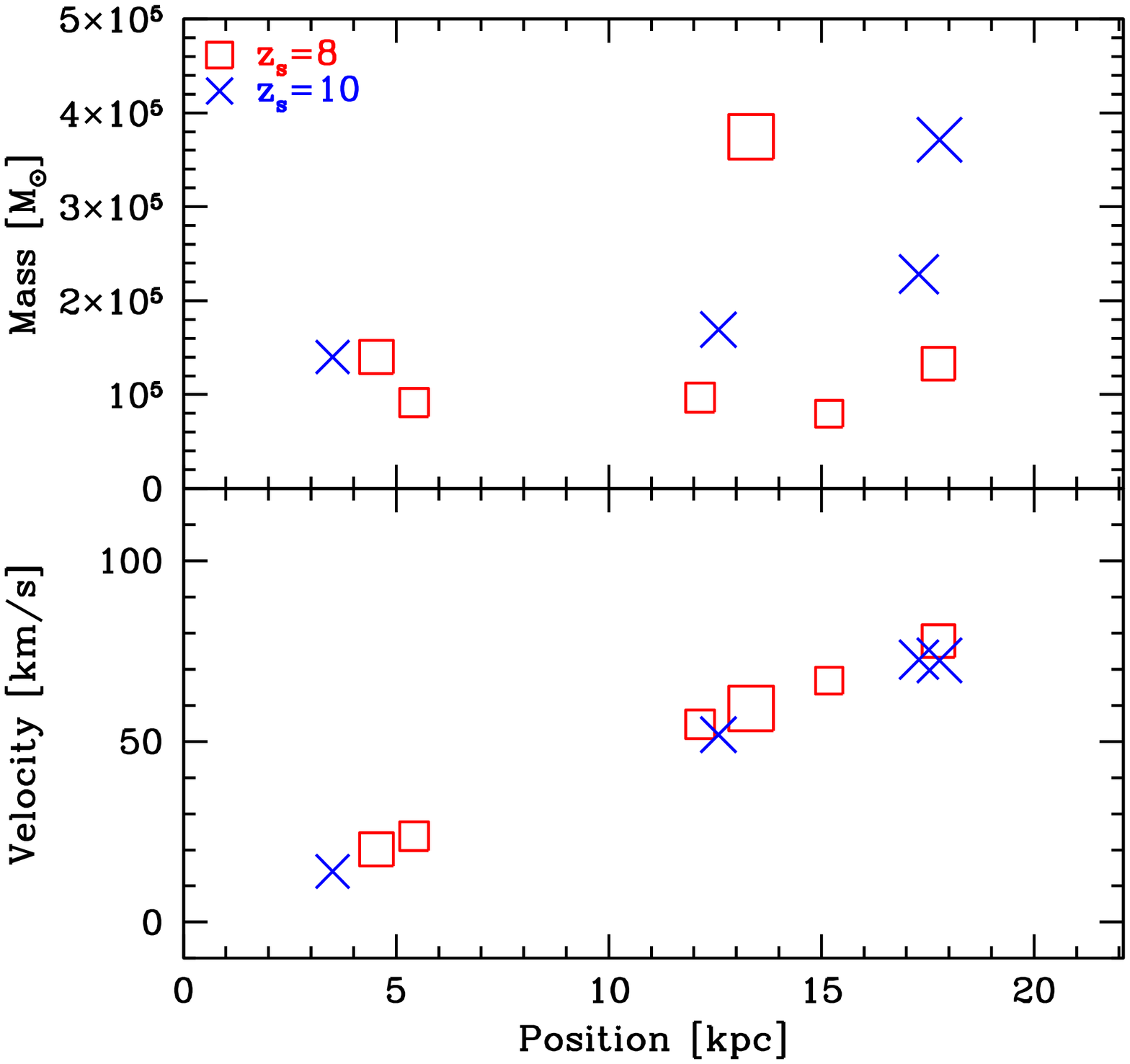}
\caption{ Comparison of  clusters generated by runs with varying outflow redshifts. The panels are the same as in Figure \ref{energy}. The (red) unfilled squares show the clusters from NFID and the (blue) crosses show the clusters from PZS10. The total mass of clusters for these models is 1.6 $\times 10^{6}$, and 2.3 $\times 10^{6}$ \msol\ for NFID and PZC10 respectively.  Although the shock lasts longer in the PZS10 case, there is little difference between these two cases. The PZS10 model has fewer, but slightly larger clusters than in the fiducial case and they are found outside of its dark matter halo. The positions and velocities are comparable between these two models. }
\label{zs}
\end{figure}

	Figure \ref{zs} compares the final stellar clusters generated by these two models. Even though the shock carries more mass in run PZS10 than in the fiducial case, the outcomes are fairly similar. Both have large clusters found outside the dark matter halo. Although fewer clusters are formed in run PZS10, the total mass in the clusters is slightly higher than in the fiducial model. 
	
\subsection{Effect of Distance from Starburst Galaxy}

In Scannapieco \etal (2004) and in Paper I and Paper II, the distance between the minihalo and the parent galaxy was taken to be $R_s$ = 1.5 $M_c^{-1/6}$$(\xi E_{55})$ $\left(\frac{1+z_c}{10}\right)^{-1}$ kpc,  a value that was chosen based on the observed metallicities of globular clusters.  While this estimate gives a fiducial value of 3.6 (physical) kpc, in this study we consider the impact of varying this key parameter over a wide range of distances between $R_s = 2.1$ kpc (PR21) and $R_s = 12$ kpc (PR120).  
	
\begin{figure*}
\centerline{\includegraphics[scale=0.5,clip, trim=50 0 0 0 ]{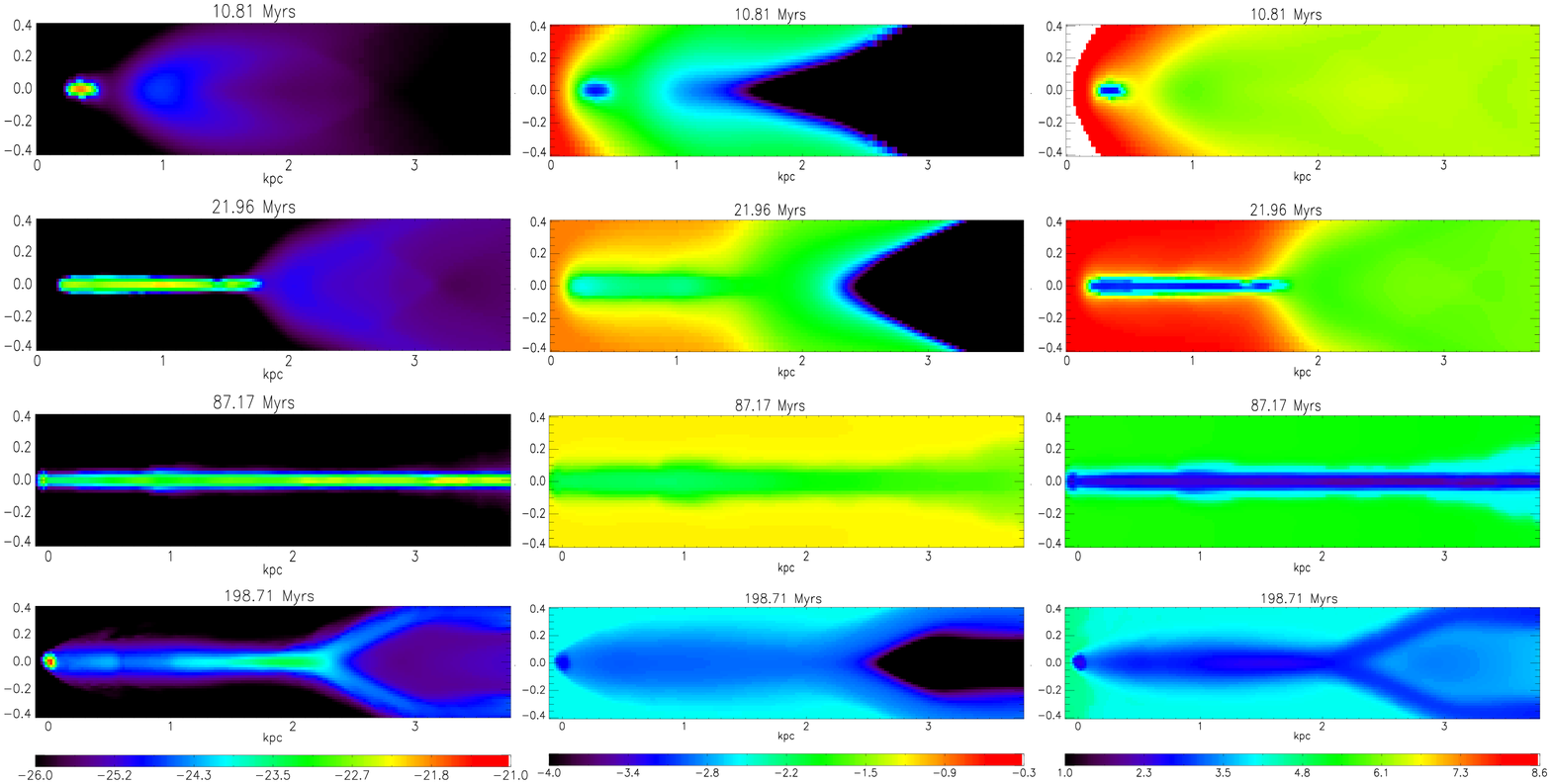}}
\caption{Comparison of final outputs from runs with varying distances between the minihalo and the galaxy. 
The first column shows logarithmic density contours between 10$^{-26}$ and 10$^{-21}$ g cm$^{-3}$, the second column shows logarithmic metallicity contours between 10$^{-4.0}$ and 10$^{-0.3}$ Z$_{\odot}$, and the third column shows logarithmic temperature contours between 10 and 10$^{8.6}$ K.
The top row shows the result of placing the minihalo at a distance of 2.1 kpc (PR21), the second is the fiducial distance of 3.6 kpc (NFID), the third row is a model with a distance of 6.6 kpc (PR66), and the last row shows the farthest case with a distance of 12.0 kpc (PR120). Beyond the apparent physical differences between each model, the metal abundance is lower than our fiducial model if the halo starts too close or too far from the outflow.  }
\label{rmerge}
\end{figure*}

	Figure \ref{rmerge} shows the final outcome from each of these models. 	When the minihalo is close to the source of the outflow, the cloud is quickly crushed into a dense cluster, similar to what occurs in runs with a larger shock energy. In fact, this happens so quickly that there is little time for the metals to be effectively mixed into this cluster and the gas ends with a lower metal abundance than our previous results. However, enough momentum is imparted to the gas to remove it from the dark matter halo.
	
	At intermediate distances (between 3.6 kpc and 6.6 kpc) the outcome is familiar. The cloud is stretched into a ribbon of gas and is expelled from the dark matter halo. The more the distance increases, the more the cloud is stretched, and there is always enough time for metals to be mixed in the primordial gas to levels consistent with our previous runs. 
	
	Finally, at the larget $R_s$ values, the cluster evolution is radically different than in the other models. The cloud is stretched as before but the resulting cluster does not leave the halo. The abundance of metals in the shock at this distance is already low, much lower than even the final cluster metallicities in many of our other runs, and while some metals are mixed into the cluster, it is very deficient compared to our other models.
	
\begin{figure}
\includegraphics[scale=0.45]{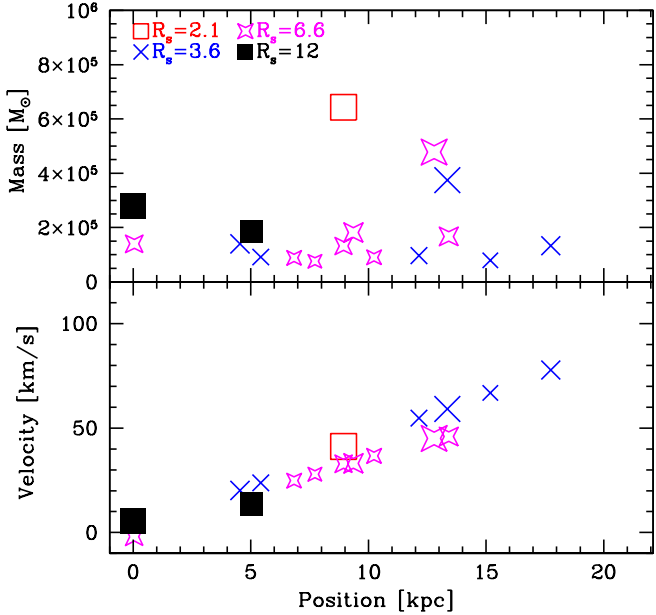}
\caption{Comparison of  clusters generated by runs with varying distances between the minihalo and the galaxy.  The (red) unfilled squares show the clusters formed in PR21, the (blue) crosses show the clusters formed in NFID, the (magenta) stars show the clusters formed in PR66, and the (black) filled squares show the clusters formed in PR120. The panels are the same as in Figure \ref{energy}.  The total mass in each model is $7.0 \times 10^{5}$, $1.6 \times 10^{6}$, $2.1 \times 10^{6}$, and $7.5 \times 10^{5}M_\odot$ for PR21, NFID, PR66, and PR120 respectively. While all models create at least one dense cluster far from the center of the halo, at the largest distances from the starbursting galaxy, the outflow is not strong enough to remove all of the gas from the halo. This leaves a cluster at the center of the dark matter halo. }
\label{distance}
\end{figure}

	Figure \ref{distance} shows the evolved state for each of these models. The panels are the same as Figure \ref{energy}. The closest halo to the galaxy is transformed into a single cluster with a total mass of $7.0 \times 10^{5}M_\odot$. The intermediate distance halos form  large clusters with masses of $\approx  3.0 \times 10^{5} M_\odot$ and several other smaller clusters with velocities and positions consistent of being free from the dark matter halo.  Finally, the farthest halo forms two dense clusters, however the largest cluster is found at the center of the dark matter halo, and only the smaller cluster is free of the halo potential at  5 kpc from its center.
	
	This suggests that there is a preferred distance from the outflow at which enriched clusters form most efficiently. Too close to the outflow and the minihalo is crushed before it is enriched and too far from the outflow and the minihalo in neither enriched nor ejected from its dark matter halo. Therefore it seems that between $\approx 3$ and 7 (physical) kpc is a preferable distance from a typical starburst for the formation of compact stellar clusters. The observed distribution of halo globular cluster positions shows a drop-off beyond a galactocentric distance of 40 kpc (Dauphole \etal 1996). After evolving these clusters, the typical distance from the galaxy is between 15-30 kpc, which agrees nicely with the observed distances.

\subsection{Effect of Halo Spin}
\label{hs}

	Next we explored the effect of net rotation on the evolution of the minihalo. In this case, the gas was given an initial velocity according to
\begin{eqnarray}
	v_x &=& -\alpha v_c \frac{y}{R_c} {\rm cos}(\phi),  \nonumber \\
	v_y &=& \alpha v_c \frac{x\ {\rm  cos}(\phi) + z\ {\rm sin}(\phi)}{R_c},  \\
	v_z &=& \alpha v_c \frac{y}{R_c} {\rm sin} (\phi),\nonumber
\end{eqnarray}
where $\alpha$ is a constant, $v_c$ is the virial velocity of the cloud, $R_c$ is the virial radius, $x$ and $y$ are the positions within $R_c$, and $\phi$ is the rotation angle. 
Here,  $\alpha$ is set at 0.05 and we vary $\phi$ between 0 and 90 degrees so that the halo rotates around the $z$-axis and $x$-axis respectively. For this value of $\alpha$ we can calculate the spin parameter of our halos using the form from Bullock \etal (2001b),
\be
	\lambda' = \frac{J}{\sqrt{2}M_{\rm vir} V_{\rm vir} R_{\rm vir}},
\ee
where J is the angular momentum of the halo with mass M$_{\rm vir}$ contained in a sphere of radius R$_{\rm vir}$ and has a circular velocity of V$_{\rm vir}$. For the halos studied this gives a spin parameter of $\lambda' = 0.023$, which is within 1 $\sigma$ of the mean value of $\lambda_0'$ = 0.035. 

\begin{figure*}
\centerline{\includegraphics[scale=0.5,clip,trim=50 0 0 0 ]{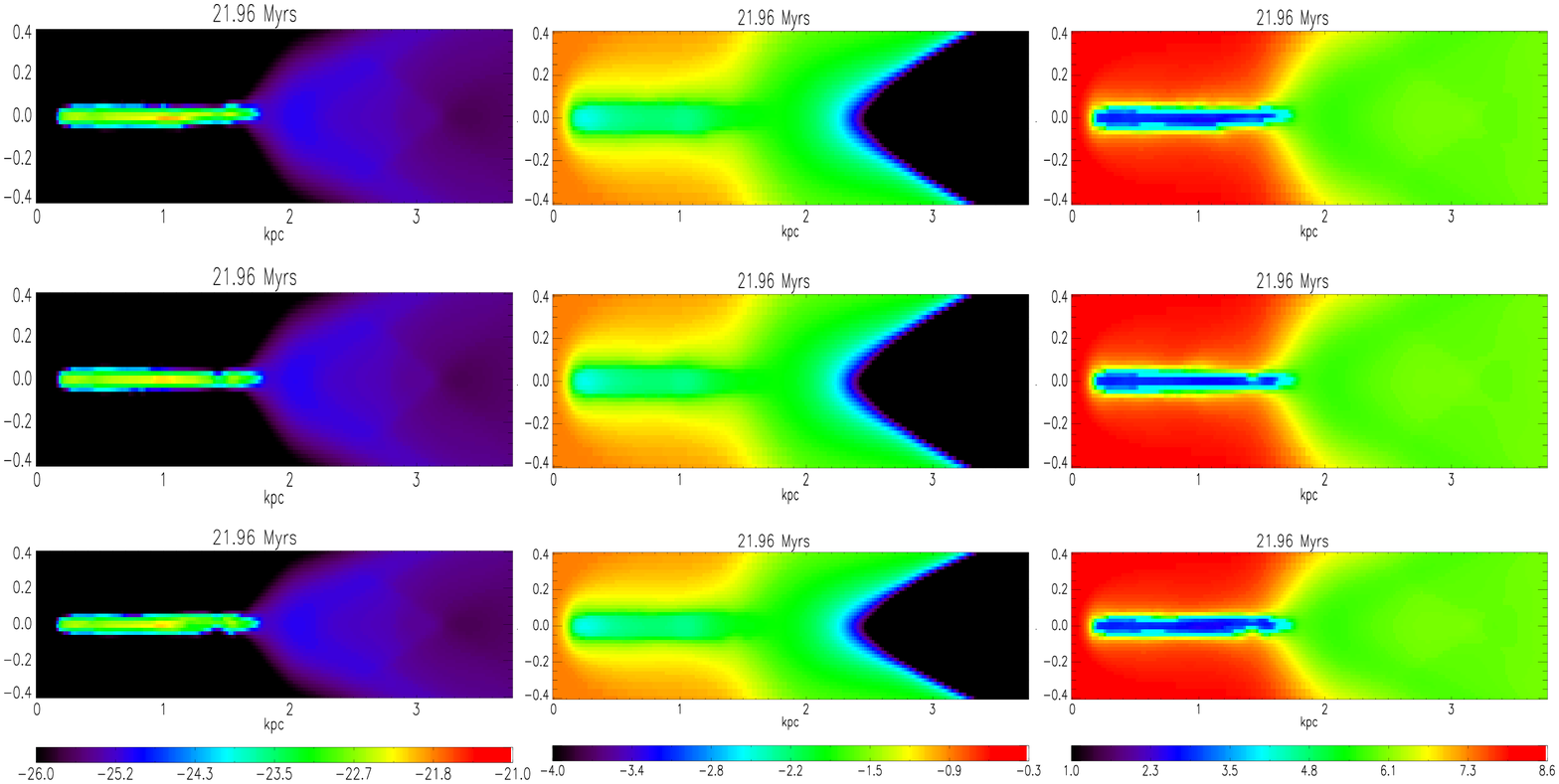}}
\caption{Final profiles of the spinning versus non-spinning halos. The top row shows the case in which the spin is about the $z$-axis (PSPZ). The middle row shows the non-rotating case (NFID) and the bottom row shows the case where the spin is about the $x$-axis (PSPN). The first column shows the logarithmic density contours, the second column shows the logarithmic metallicity contours, and the third column shows the logarithmic temperature contours. Panel limits are the same as in Figure \ref{mmass}. There is very little difference between each of these runs, which suggests that the initial spin of the minihalo does not contribute significantly to its final evolution. }
\label{smerge}
\end{figure*}

Figure \ref{smerge} compares the simulations at the final time. It is obvious that neither the spin direction nor the magnitude of the spin changes the final distribution. In all cases the cloud is stretched into a ribbon along the $x$-axis, the primordial gas is enriched to nearly constant value near 10$^{-2.0}$ Z$_{\odot},$ and the cluster is cooled to a few hundred degrees K. 

The reason for this insensitivity to $\lambda'$ is a result of the shock itself. As the shock advances toward the halo it begins to develop vorticity, defined as $\vec{\omega} \equiv \vec{\triangledown} \times \vec{v},$ which evolves as $\frac{D\vec{\omega}}{Dt} \approx \frac{1}{\rho^2} \vec{\triangledown} \rho \times \vec{\triangledown} p $, where $v$ is the velocity, $\rho$ is the density, and $p$ is the pressure. Physically, this baroclinic source term is a measure of the generation of vorticity due to the mismatch between the gradients of pressure and density (Glasner \etal 1997). Since the shock is not aligned with the density gradient of the halo, this term is large throughout the simulation and vorticity begins to grow rapidly. 

\begin{figure}
\includegraphics[scale=0.40,clip,trim=50 0 0 0 ]{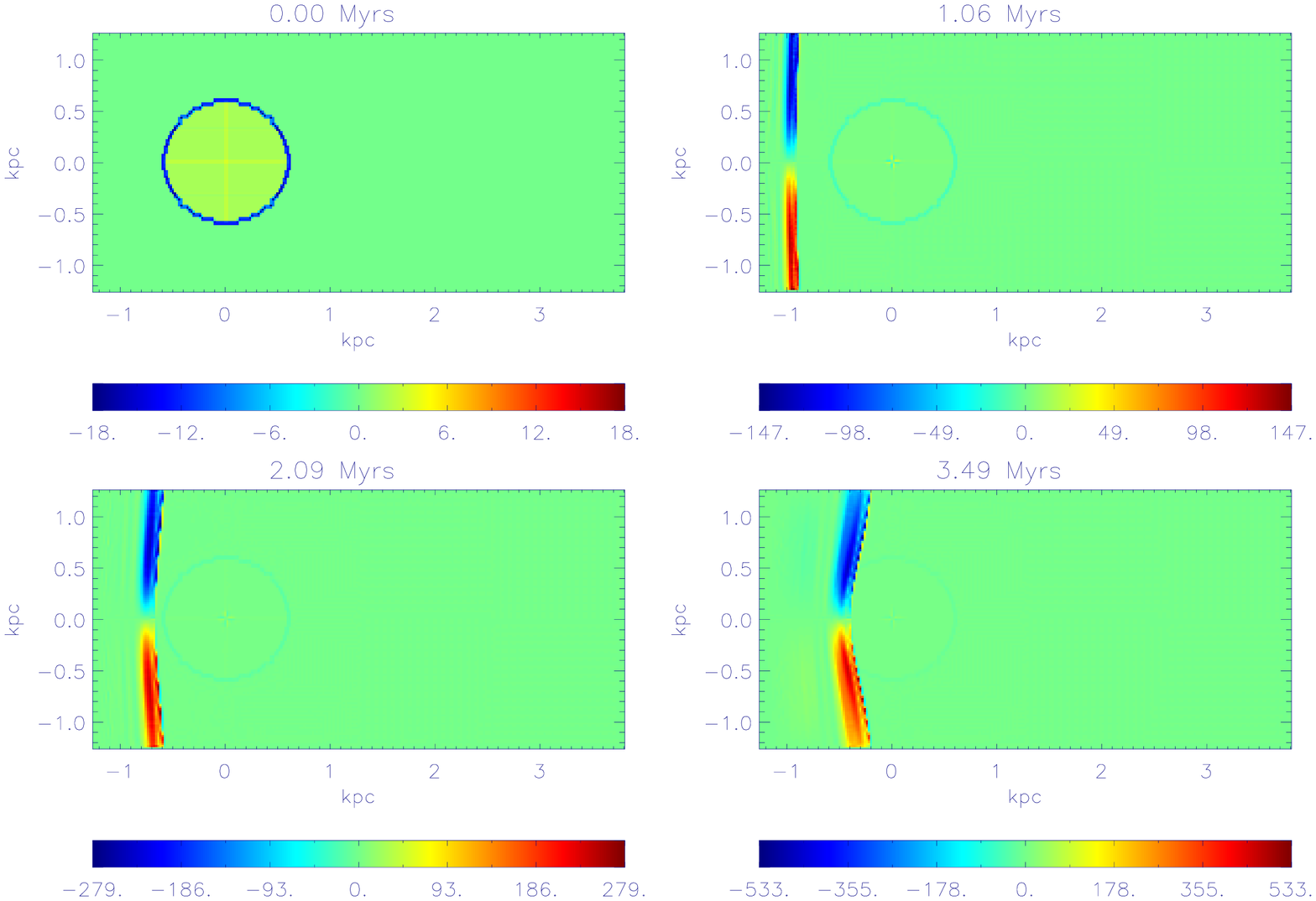}
\caption{Snapshots of the $z$-component of vorticity at different times before the shock impacts the halo in run PSPZ. The top left panel shows the initial vorticity of the cloud. The top right and both bottom panels show the evolution of vorticity as the shock nears the halo. The scale is in units of km/s/kpc. As the shock nears the minihalo the vorticity increases and by the time it reaches the halo, it is much larger than the spin of the halo. }
\label{vort}
\end{figure}

Figure \ref{vort} shows the $z$-component of vorticity as the shock heads toward the halo.  By the time they meet the magnitude of the vorticity is much greater than the spin of the halo. In fact, the vorticity would be much greater than the spin of the halo even if we had chosen an $\alpha$ of 1.0.

\subsection{Effect of Halo Concentration}
	As the name suggests, the concentration parameter, $c$, describes in Eq.\ref{enfw} how the density is aggregated toward the center of the halo. Less concentrated halos have lower maximum central densities and shallower density profiles, while more concentrated halos have higher central gas densities and steeper density profiles at large radii. It is defined as 
\be
	c \equiv R_{200}/R_s,
\ee
where R$_{200}$ is the radius that corresponds to a density that is 200 times the critical density (Navarro \etal 1997) and $R_s$ is the inner radius of the cloud. Typical values for NFW halos at this redshift are 4.8 (Madau \etal 2001). However, Bullock \etal (2001a) used high-resolution N-body simulations to study the dark matter halo density distribution and found that for a given mass the range in concentration parameter can be fairly large, 1$\sigma$ $\Delta({\rm log\  c}) = 0.18$.  We therefore studied interactions with halos that cover this spread in concentration.

\begin{figure*}
\centerline{\includegraphics[scale=0.5,clip,trim=50 0 0 0 ]{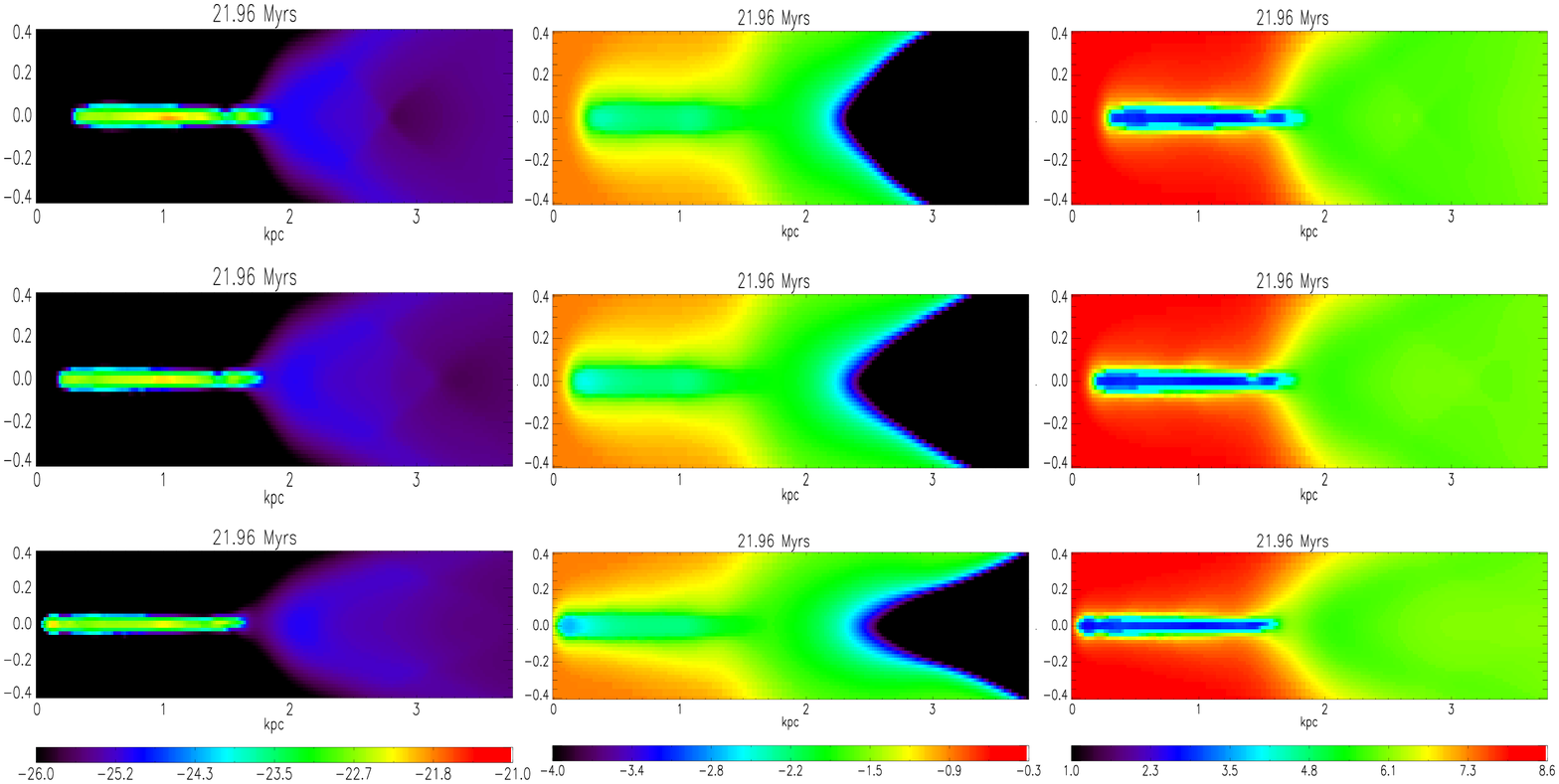}}
\caption{Comparison of final outputs from runs with different concentration parameters. The first row shows the $c=3.2$ case (PC32), the second row shows the fiducial $c=4.8$ case (NFID), and the third row shows the $c=7.3$ case (PC72).  Columns are the same as Figure \ref{mmass}. The more concentrated the halo,  the more stretched it becomes and the less it is enriched.}
\label{concenmerge}
\end{figure*}

Figure \ref{concenmerge} compares models of the minihalo after altering the concentration parameter.  The top, middle, and bottom rows present the results of runs with $c=3.2$, 4.8, and 7.3 respectively. As the halo becomes more concentrated, the metals have a harder time enriching the center of the halo, and becomes harder for the gas to be pushed out of the halo, which can be seen in the first column where the halo with the lowest concentration is farther away from the center of the halo.

\begin{figure}
\includegraphics[scale=0.40]{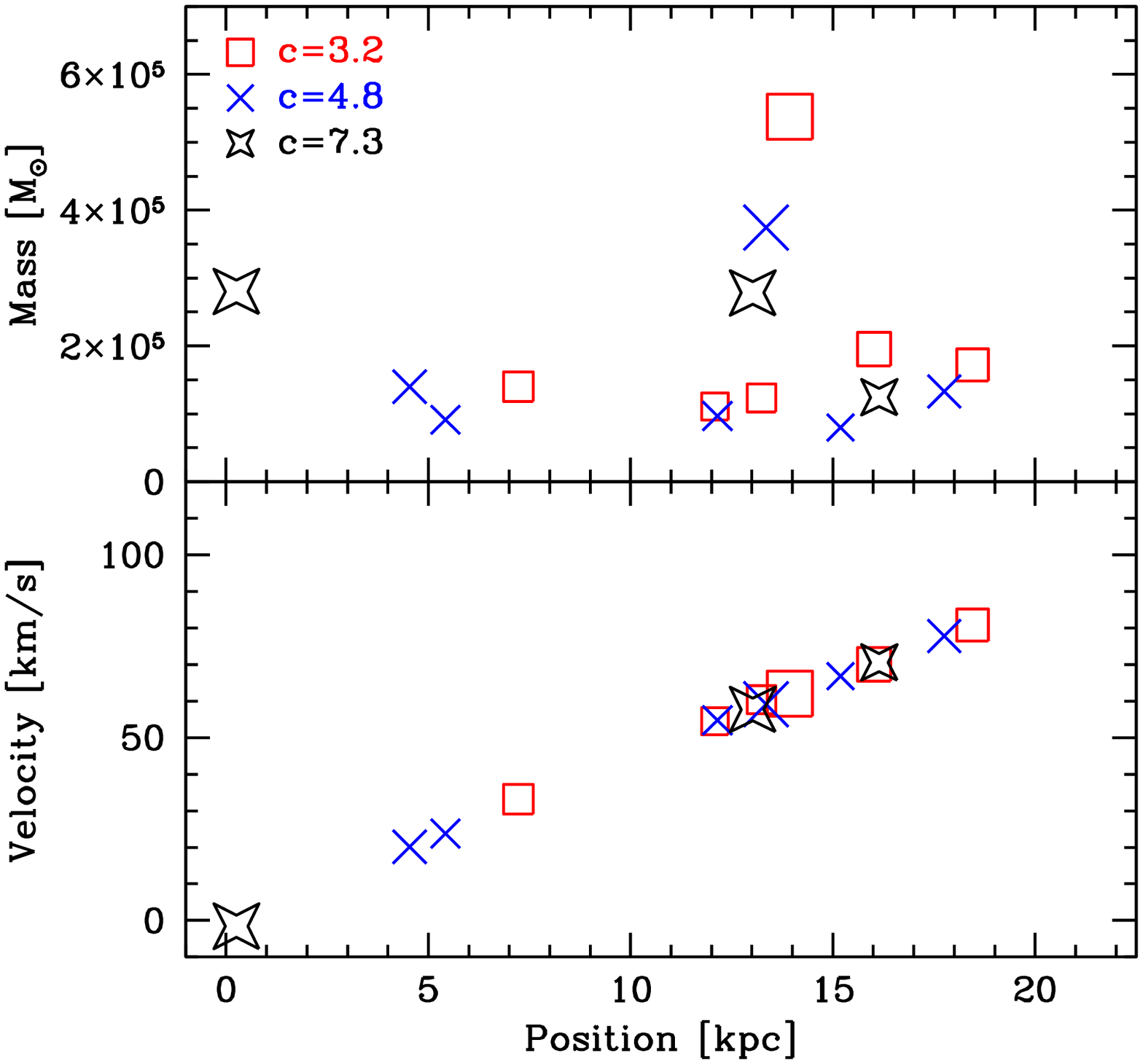}
\caption{Comparison of clusters generated in runs with different halo concentrations. Panels are the same as Figure \ref{energy}.  The (red) unfilled squares show clusters from the least concentrated halo (PC32), the (blue) crosses show the clusters from the fiducial cluster (NFID), and the (black) stars show the clusters from the most concentrated halo (PC73). The total mass of clusters in each model is 1.8 $\times 10^{6}$, 1.6 $\times 10^{6}$, and 1.5 $\times 10^{6} M_\odot$ for PC32, NFID, and PC73 respectively.  The outflow has a hard time removing the gas from the most highly concentrated halo, and in this run a cluster is formed within the dark matter potential well.  On the other hand, all the gas is ejected from the least concentrated halo, which forms the most massive cluster of any of the three runs.}
\label{concen}
\end{figure}

Figure \ref{concen} shows the long term evolution of these models. Generally there is little difference between these models, which all contain at least one large cluster with a mass few $\times 10^{5} M_\odot$. However, since there is more gas at near the edge of the less concentrated halos, the total mass of the ribbon is greater. Finally, in the most concentrated halo there is a portion of the gas that fails to leave the dark matter halo. 

\subsection{Summary}

Table~\ref{sumt} shows the final outcomes from all of our runs, including the total mass of clusters formed, and the basic properties of the largest cluster. Our findings can be summarized as follows:

\begin{enumerate}

\item {{\em The mass of the initial minihalo is closely correlated with the final mass of the compact clusters,} such that the more massive the initial halo, the more massive the total distribution of clusters.   However, as the halo gets more massive, not all the gas is removed from the halo, and a sizable cluster forms at the center of the dark matter potential. In all cases the gas is enriched to nearly identical metal abundances, and only the center of the most massive minihalo is slightly less enriched. }

\item {{\em The total energy in the outflow primarily affects the degree to which the final distribution is stretched.} At low energies, the gas is shaped into a  long ribbon of material that is enriched at a lower level than the fiducial model. As the energy increases, the ribbon becomes smaller until the halo is completely crushed, forming a single dense cluster with a similar metallicity as in the fiducial case. }

\item {{\em The collapse redshift of the minihalo directly affects the number and distribution of collapsed clusters.} Halos that collapse early each produce a massive cluster that is found far from the dark matter halo. Halos that collapse later form a large number of lower mass clusters. Generally the clusters' metal abundances are independent of minihalo collapse redshift, but the earlier redshift runs yield clusters with slightly metal-deficient cores.}

\item{{\em If the redshift of the shock is increased, it entrains more mass, moves slower, and stretches the minihalo gas more efficiently.}  At lower redshifts, the shock crushes the cloud more efficiently, and a more compact ribbon is formed. At higher redshifts, the cloud is enriched slightly more than the lower redshift cloud, however both have an average abundance of $\approx 10^{-2}$ Z$_{\odot}$. }

\item{{\em The distance between the starburst galaxy and the minihalo has a dramatic impact on all properties of the final distribution of clusters.}  At small separations the minihalo is quickly crushed and is deficient in metals. At slightly larger distances, metals are found at levels similar to our fiducial model, and the cloud is stretched into a ribbon of material, which becomes longer in runs with larger distances. At very large distances the cloud is disrupted but the outflow is too weak to move much of the gas out from the dark matter halo. Because the shock itself is metal-poor, the final cluster is also metal-poor.  }

\item{{\em The initial spin of the halo has no discernible impact on the evolution of the minihalo.}  This is because the vorticity generated during the shock minihalo interaction dwarfs that of the minihalo by orders of magnitude. All cases studied were identical to the case without spin. }

\item{{\em The concentration of the halo has a strong effect on the positions of the collapsed stellar clusters.}  The more concentrated the minihalo, the harder it is for the outflow to remove the gas from the dark matter potential, and  in the most concentrated  case we studied, some halo gas remains in the dark matter halo. In all cases the gas is enriched to nearly identical levels with only the most concentrated gas at the center of the halo being slightly metal deficient. However, dense clusters are also always formed outside of the dark matter potential.} 

\end{enumerate}

\begin{table*}
\begin{center}
\caption{ Summary of model outcomes. $\rm{M_{halo}}$ is the initial gas mass of the minihalo in units of 10$^{6}$ M$_{\odot}$, M$_{ {\rm clusters}}$ is the total mass found in clusters in units of 10$^{6}$ M$_{\odot}$, M$_{{\rm large}}$ is the mass of the largest cluster formed in units of 10$^{6}$ M$_{\odot}$, Z$_{\rm cluster}$ is the metallicity of the largest cluster, V$_{ {\rm cluster}}$ is the velocity of the cluster after 200 Myrs of evolution in units of km s$^{-1}$, and D$_{ {\rm cluster} }$ is the distance of the largest cluster from the center of its dark matter halo in units of (physical) kpc.   }
\begin{tabular}{cccccccc}
\hline
 Name           & M$_{ {\rm halo}}$               & M$_{ {\rm clusters }}$           & M$_{ {\rm large }}$              & Z$_{{\rm cluster}}$  & V$_{ {\rm cluster}}$ & D$_{{\rm cluster}}$ & Notes \\
                       &$10^{6} \ {\rm M_{\odot}}$  & $ 10^{6} \ {\rm M_{\odot}}$  & $ 10^{6} \ {\rm M_{\odot}}$ & {\rm Z$_{\odot}$}      &  km s$^{-1}$   &   kpc                 &             \\
 \hline
 OFID	     & 0.45 			      & 0.40				& 0.07			  &$10^{-2.0}$	     & 45	                & 9.8                &            \\
 PM10/NFID & 1.5 			      & 1.6				& 0.38			  &$10^{-2.0}$	     & 59	                & 13	       &	  \\
 PM03	    & 0.05 		               & 0.01			         & 0.01			  &$10^{-2.0}$       & 76		      & 16	       &	  \\
 PM30	    & 4.5 		 	      & 4.8				& 1.9			           & $10^{-2.0}$      & 63	 	      & 15	       &	  \\
 PE1		    & 1.5 		 	      & 0.9				& 2.8			           & $10^{-3.0}$      & 17	 	      & 5.0	       & Data for cluster outside of DM halo	  \\
 PE5		    & 1.5 			      & 2.3				& 0.37			  & $10^{-2.5}$      & 72		      & 17	       &           \\
 PE20	    & 1.5 			      &	0.9				& 0.66			  &$10^{-2.2}$	     & 41                     & 9.6                &           \\
 PE30	    & 1.5			      & 0.8				& 0.82			  & $10^{-2.0}$      & 53		      &	12	       &	  \\
 PZC8	    & 1.5			      & 1.2				& 0.32			  & $10^{-2.0}$      & 60		      & 13                &           \\
 PZC15         & 1.5			      & 2.2				& 0.96			  & $10^{-2.0}$      & 130      	      & 29                &		  \\
 PZS10	    & 1.5			      & 2.3				& 0.37			  &  $10^{-1.5}$     & 72		      & 18	       &           \\
 PR21	    & 1.5			      & 0.7				& 0.64			  & $10^{-3.0}$      & 41		      & 8.9	       &	  \\
 PR66 	    & 1.5			      & 2.1				& 0.48			  & $10^{-2.0}$      & 45		      & 13	       &           \\
 PR120         & 1.5			      & 0.75				& 0.19			  & $10^{-3.0}$      & 14		      & 5.0		       & Data for cluster outside of DM halo	  \\		
 PC32           & 1.5			      & 1.8				& 0.54			  & $10^{-2.0}$      & 62		      & 14               &           \\
 PC73	    & 1.5			      & 1.5				& 0.28			  & $10^{-2.0}$      & 58		      & 13	       & Data for cluster outside of DM halo \\
 \label{sumt}
\end{tabular}
\end{center}
\end{table*}

\section{Observational Signatures}

The suite of simulations described above shows that over an extremely wide range of parameters, the ultimate consequence of the interaction between a galactic outflow and a primordial minihalo  is the rapid formation of dense clusters containing up to  a few $10^6 M_\odot$ of stars.   While such high-redshift clusters are not directly observable with current telescopes, their rapid bursts of star formation and consequent  low mass-to-light ratios present a opportunity for study with the next generation of instruments.   At the same time, their compact nature and formation in low-density environments  makes it likely that many of them may have survived to the present,  allowing for indirect connections with current stellar populations. Here we explore both of these connections.

\subsection{Direct Observations}

       To calculate the direct observability of the clusters formed in our simulations, we first constructed an estimate of the number of stars formed as a function and time and position.  As there is no explicit prescription for star formation included in our simulations, we instead built up the star-formation history by post-processing our outputs,  carrying out the following steps:

\begin{enumerate}
\item First we calculated the total stellar mass in a series of 175 evenly-spaced  bins,  adding together the mass in cells that exceeded a density threshold of $\rho_{\rm th} = 1.0 \times 10^{-23}$ g cm$^{-3}$.  Note that as the mass collapses dramatically onto the $x$ axis, its final density is limited by the resolution in the $y$ and $z$ directions.  Thus the threshold is less than the density of the collapsed cloud in nature, and it was determined by examining the late stages of our fiducial model and finding the apparent edge of the ribbon of material.  In this way the collapsed mass was computed for every simulation output, such that we computed the masses and position of the stars as a function of time throughout the simulation. It should be noted that the masses quoted below are assume that the star formation efficiency is 100 percent, and therefore the quoted stellar mass and stellar flux are an upper limits.  

\item  Next, we used the positions and velocities to correlate the stellar distribution at each output with the distribution at the previous output. Starting with with the final output,  and working back through the files in this way,  we calculated the mass of new stars formed as a function of time and determined  the positions of these stars at the final output time.

\item Finally, we calculated the stellar fraction as a function of age for each of the mass bins at the final output time.

\end{enumerate}

      Using this information, we could then estimate the emitted flux for an arbitrary set of  narrow or wide band filters.   At redshifts $z \geq 8$ the {\it James Webb Space Telescope (JWST)} will be the best telescope to see these objects in wide band filters. From the ground, the best method of detection is to search for redshifted Lyman alpha (Ly${\alpha}$) emission  using near-infrared capabilities of next generation, 30-40 meter class telescopes such as the {\em  Giant Magellan Telescope (GMT)}, the  {\em Thirty Meter Telescope (TMT)}, and the {\em European Extremely Large Telescope (E-ELT)}.

To determine the broad-band fluxes of our stellar clusters, we used the population-synthesis code {\rm bc03} (Bruzual \& Charlot 2003) to compute the luminosity per solar mass of a stellar population as a function of frequency and age,  and convolved this with the stellar history of each bin, and related this to the flux at an observed frequency $\nu_0$ as
\be
F_{\nu_0} = \frac{(1+z)}{4 \pi d_l^2(z)} \int dt \frac{d L_\nu}{d M_*} \left(\frac{\nu_0}{1+z},t \right) \frac{d M_{*}}{dt}(t), 
\ee
in units of ergs cm$^{-1}$ Hz$^{-1}$ s$^{-1}$ and where $z$ is the shock redshift, $d_l(z)$ is the luminosity distance, $\nu_0/(1+z)$ is the rest-frame frequency, $c$  and $d L\nu /dM_* (\nu_0/(1+z),t)$ is the luminosity per frequency per solar mass of a population of stars with an age $t$, and $dM_{*}/dt(t)$ is the star formation history in a given bin. 

Similarly, we estimate the Lyman alpha flux from our simulations as
\be
       F_\alpha = \frac{1}{4 \pi d_l^2(z)}   \int dt \frac{d L_\alpha}{d M_*}(t)  \frac{d M_{*}}{dt}(t),
\ee
where $d L_{\alpha}/d M_\star (t) = c_L (1-f_{\rm esc}) Q(H) M_{*},$ is the  the Lyman alpha luminosity per solar mass of a population of stars with an age $t$  (\eg Scannapieco \etal 2003), with $c_L$ $\equiv$ 1.04 $\times 10^{-11}$ ergs, $f_{\rm esc}$ is the escape fraction of ionizing photons and is taken to be 0.2,  $Q(H)$ is the hydrogen ionizing photon rate in units of number per second per solar mass and is taken from  the stellar population synthesis code STARBURST99 (Leitherer \etal1999). Finally, in both the broad and narrow band cases we normalized the flux by the spacing of each bin to get the flux per (physical) kpc.

\begin{figure*}
\begin{center}
\includegraphics[scale=0.80,clip,trim=0 0 0 0 ]{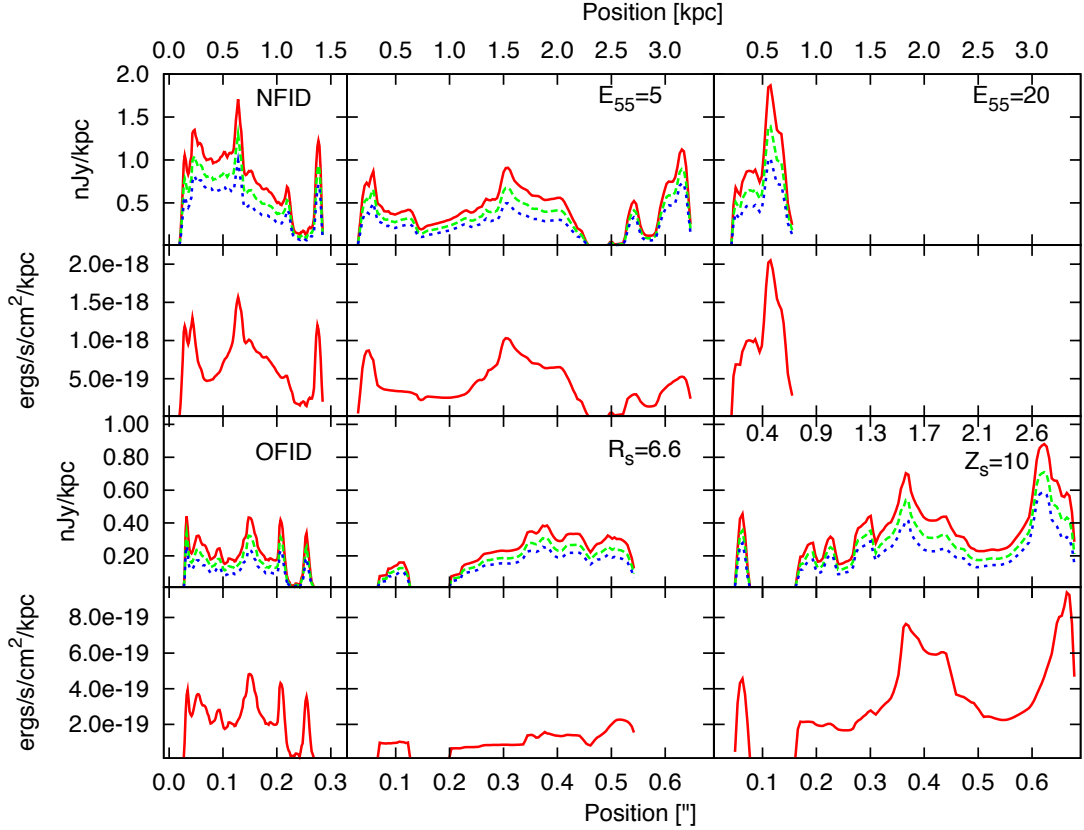}
\caption{Simulated fluxes from selected models, scaled to 100\% star formation efficiency. The top two rows show the simulated fluxes for runs NFID, PE5, and PE20 (from left to right), while the bottom two rows shows the simulated fluxes for runs OFID, PR66, and PZS10. The first and third rows show the expected fluxes in {\it JWST} wide band NIRCam filters. The (red) solid lines are fluxes in the F115W band, the (dashed) green lines are in the F150W band, and the (dotted) blue lines are in the F200W band, all measured in nJy per kpc. The second and fourth rows show the expected Ly${\alpha}$ intensities and are measured in ergs/s/cm$^2$/kpc. The top $x$-axis is the physical spatial scale measured in kpc of each model while the bottom $x$-axis is the angular scale measured in arcseconds. The $y$-axis is identical across a given row.  }
\label{jwstflux_fid}
\end{center}
\end{figure*}

Figure \ref{jwstflux_fid} shows the resulting fluxes for three different {\it JWST} bands (F115W, F150W, and F200W) and the Lyman alpha flux from a selection of simulations spanning a wide range of parameter space.   Again we note that the fluxes in this figure are computed assuming that all the gas collapsed above $\rho_{\rm th} = 1.0 \times 10^{-23}$ g cm$^{-3}$  is converted into stars, and so these fluxes should be scaled by an unknown star formation effciency factor which is $\leq 1.$ Likewise the angular given in this figure assume that that stellar clusters are being viewed edge on, such that angular separation is maximal.  Table~\ref{msf} summarizes the observable properties of the stellar clusters generated in all of  our models including the total fluxes in the {\it JWST} bands and Ly${\alpha}$ and the physical and maximal angular scales. Note that the physical and angular scales given are for the extended emission expected at the end of each simulation and not from the final globular cluster.  

From the values in this figure and table we see that, unfortunately even the wide band filters of {\em JWST}  are not expected to have the required sensitivity to detect these objects. Observations with {\em JWST}  wide band filters will have typical sensitivities of 10-20 nJy for a 10-$\sigma$ detection at 10,000 seconds of integration time (Stiavelli \etal 2008),  which is roughly an order of magnitude higher than the typical fluxes of $\approx$ 1 nJy (see Figure \ref{jwstflux_fid}) expected from our objects.

On the other hand, the narrow band Ly${\alpha}$ fluxes are over an order of magnitude {\em brighter} than those expected to be obtained with the next generation of ground-based telescopes.
The proposed {\em Near Infrared Multi-object Spectrograph (NIRMOS)} on the {\em GMT} for example,  will be able to detect such sources down to a flux limit of 1.0$\times 10^{-20}$ ergs/s/cm$^{-2},$ given 25 hours of integration time (McCarthy 2008; GMT Science Case). Similarly, the {\em InfraRed Imaging Spectrometer (IRIS)} on the {\em TMT} will detect Ly${\alpha}$ sources at z = 7.7 with fluxes of 1.0$\times 10^{-18}$ ergs/s/cm$^{-2}$ with signal-to-noise (S/N) of 15 in only 1 hour of observation time (Wright \& Barton 2009; TMT Instrumentation and Performance Handbook). Finally, {\em OPTIMOS-EVE (Optical-Near-Infrared Multi-object Spectrograph)} for the {\em E-ELT} will detect sources with fluxes of 10$^{-19}$ erg/s/cm$^{-2}$ with S/N of 8 in 40 hours of integration time (Hammer \etal 2010).   This means that the majority of the models studied here are more than bright enough to be detected.

\begin{figure*}
\begin{center}
\includegraphics[scale=1.30,clip,trim=0 0 0 0 ]{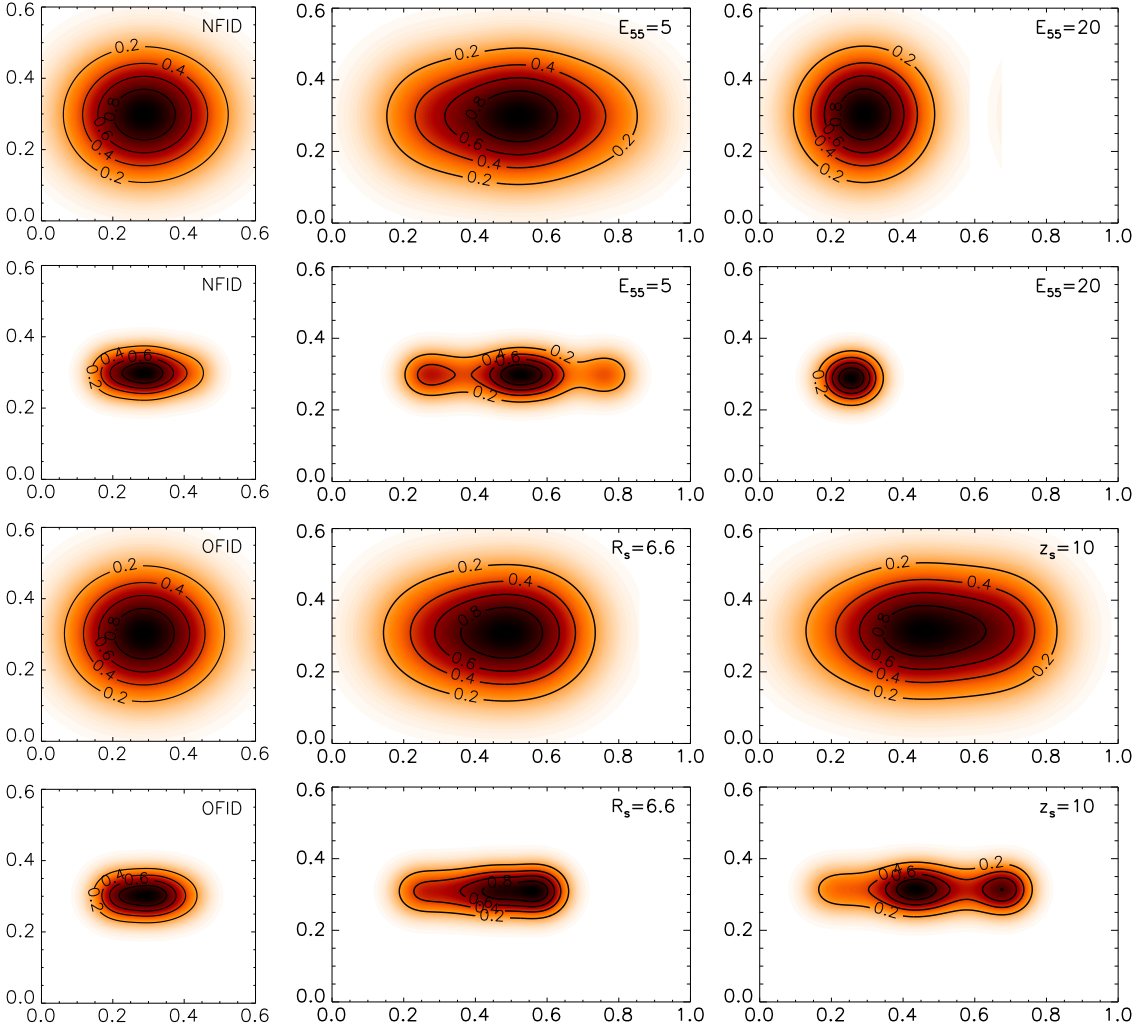}
\caption{Simulated narrow-band images from selected models. The top two rows show simulated images for runs NFID, PE5, and PE20 (from left to right), while the bottom two rows shows the simulated images for runs OFID, PR66, and PZS10. The first and third rows show forming clusters as observed edge-on with 0.25 arcsecond resolution, and the second and fourth rows show clusters as observed edge-on with 0.1 arcsecond resolution.  In each panel the $x$ and $y$ axes are in units of arcsec, and the  contours are labeled according to flux per unit area relative to the maximum flux per unit area in the image.}
\label{narrowband_image}
\end{center}
\end{figure*}

Beyond being bright in Ly$\alpha$, the unusual, elongated morphology of the stellar distributions formed in outflow-minihalo interactions
makes them ideal for study with next generation  of ground-based instruments.   With expected angular resolutions of $0.1-0.3$ arcsec (McCarthy 2008; Wright \& Barton 2009; Hammer \etal 2010), next generation narrow band images will  not only be able to detect the presence of the stars, but show that their distribution is highly-elongated in the direction of the impinging outflow, as illustrated in Figure \ref{narrowband_image}. 
Furthermore, as the starburst galaxies triggering star-formation should have typical solar masses $\gtrsim 10^8 M_\odot$, these will be easily detectable with broad-band {\em JWST} measurements.    Thus, the detection of a group of elongated Ly$\alpha$ emitters which whose axes point directly at a larger broad-band detected starbursting galaxy will provide a unique signature that  unambiguously points to the formation of compact stellar clusters by high-redshift galaxy outflows.


\begin{table}
\begin{center}
\caption{Summary of simulated model fluxes. F115W is the total flux from the F115W {\it JWST} band in units of nJy, Ly$_{\alpha}$ is the total Lyman $\alpha$ flux in units of ergs/s/cm$^2$, and the third and fourth columns are the maximum angular and physical spatial scales respectively.}
\begin{tabular}{ccccc}
\hline
     Model &        F115W & Ly$_{\alpha}$ & arcsec ('') &        kpc \\
\hline
      OFID &       0.25 &   9.2e-19 &       0.27 &       1.3 \\
      NFID &       1.02 &   2.8e-18 &       0.28 &        1.4 \\
      PM03 &      0.02 &  2.3e-20 &       0.14 &        0.7 \\
      PM30 &      2.7   &  3.0e-18 &       0.55 &        2.7 \\
       PE5 &        1.3 &   1.2e-18 &       0.65 &        3.2 \\
      PE20 &        0.54 &   6.1e-19 &       0.16 &       0.76 \\
      PE30 &        0.40 &   4.6e-19 &       0.08 &       0.39 \\
      PZC8 &       0.53 &   5.4e-19 &       0.24 &        1.2 \\
     PZC15 &          2.1 &   2.3e-18 &       0.46 &        2.3 \\
     PZS10 &       0.89 &   8.9e-19 &       0.68 &        2.9 \\
      PR21 &        0.35 &   4.0e-19 &       0.05 &       0.25 \\
      PR66 &       0.47 &   2.3e-19 &       0.54 &        2.7 \\
      PC32 &       0.99 &   8.5e-19 &       0.27 &       1.3 \\
      PC72 &       0.68 &   7.2e-19 &        0.30 &       1.5 \\
\label{msf}
\end{tabular} 
\end{center}
\end{table}

\subsection{Connection with Halo Globular Clusters}

Equally exciting as the prospect of direct detection is the connection between  outflow-induced star formation at high-redshift and  stellar clusters observed around nearly galaxies.
Over a large range of parameter space, the interactions studied in our simulations result in compact, tighty-bound stellar clusters that are found in low-density environments in which cluster disruption processes are minimal.
Thus it is likely that a substantial fraction of the clusters generated by outflow-minihalo interactions may persist even to the present day.  As mentioned above and in our previous investigations (Paper I and Paper II), a natural low-redshift counterpart of these compact, high-redshift  clusters may be the population of  halo globular clusters.

How globular clusters are formed has been an open question since their discovery. The typical age of a globular cluster is between 10-13 Gyrs (Krauss \& Chaboyer 2003), which demands a cosmological origin for these systems. Observations of galactic globular clusters have found a bimodal color distribution which suggests two subpopulations (Zepf \& Ashman 1993).  The relatively metal rich population with $\left[\frac{Fe}{H}\right] \approx -0.5$ are associated with the thick disk or bulge of galaxies and are thought to have formed as a consequence of galaxy interactions during the build-up of their parent galaxy (Shapiro \etal 2010; and see Brodie \& Strader 2006 for a review).  On the other hand, the metal-poor population with $\left[\frac{Fe}{H}\right] \approx -1.6$ are associated with galaxy halos and  are thought to form in the early universe through an unknown mechanism.

	Formation scenarios for such clusters are typically split into two groups: `pre-enrichment' and `self-enrichment' schemes. In the pre-enrichment picture, the primordial gas is homogeneously enriched via supernovae in such a way that does not disrupt the cloud (\eg Beasely \etal 2003;  Bromm \& Clarke 2002; Elmegreen \& Efremov 1997; Marcolini \etal 2009). However, little is known about this previous generation of stars, how it was able to enrich the gas so quickly, and why it played only a secondary role in the formation of globular clusters.  In the self-enrichment picture the primordial gas was enriched by a supernova contained within the halo hosting the forming globular cluster (\eg Boley \etal 2009; Brown \etal 1995; Recchi \etal 2005; Smith 2010). This has the problem of not being able to mix the metals into the cluster rapidly enough, as well as usually leads to the the supernova unbinding the cluster (Peng \& Weisheit 1991; Whalen \etal 2008b). Finally, Cen (2001) set aside the question of enrichment completely and used a simple model to suggest  that ionization fronts can act as an external force to collapse these clouds.  However, many others (\eg Haiman \etal 2001; Illiev \etal 2005; Shapiro \etal 2004) showed that instead of forming a dense cluster, the ionization front completely boils away the minihalo.   As shown in our simulations,  interactions between galaxy outflows and minihalos share the best aspects of all these scenarios: bringing in metals only moments before vigorous star formation commences, depositing the metals without unbinding the proto-globular cluster, and triggering collapse without evaporating the cloud.	 
	
       Observationally, there are three important properties of halo globular clusters that make this connection a promising one.  First, globular cluster masses are distributed as a Gaussian with a mean value of 10$^{5} M_\odot$  and a dispersion of 0.5 dex (\eg Armandroff 1989).  Here, the lower mass cut-off in this population is likely to be due to a variety of destruction processes including mechanical evaporation (\eg Spitzer \& Thuan 1972) and shocking as the cluster moves through the disk of the galaxy (\eg Ostriker \etal 1972), but the high-mass cut-off appears to be a property of the initial population. Except for the models with the smallest minihalo masses, all of our models produce at least one dense cluster with a mass between 10$^{5}$ and a few times 10$^{6}$ \msol. Furthermore minihalos have an intrinsic maximum mass of  $\approx 10^{7} \msol$ which corresponds to the T $\approx 10^{4}$ K limit where atomic hydrogen/helium cooling becomes inefficient, and thus the maximum sizes of compact clusters in our study are likely to place a rough upper bound on the masses of stellar clusters that can be formed by this mechanism in nature.

       A second important property connecting our high-redshift clusters with the present-day population of halo globuar clusters is the abundances of stars  both within a given cluster and between different halo globular clusters. The metallicity distribution between clusters is well defined by a Gaussian distribution with a mean value $\left[\frac{Fe}{H}\right] \approx -1.6$ with a dispersion of 0.3 dex (Zinn 1985; Ashman \& Bird 1993). Most individual clusters have a dispersion of less than 0.1 dex (see Suntzeff 1993 and references within), although it is worth noting that some of the clusters that show larger scatter may be due to subsequent star formation and enrichment due to evolved stars (\eg Piotto \etal 2007; D'Ercole \etal 2008; Bekki 2011).

     In our simulations,  the metals from the incoming outflows are well mixed into the primordial gas from the minihalo through the turbulent processes that are inherent to this interaction.   Over a wide range of parameters, the metal-free gas is enriched to nearly constant value of Z $\approx$ 10$^{-2}$ Z$_{\odot}$, which approximately matches the observations. Only a small subset of models are found where the metal abundance is below this value. In cases where the halo is very close to the galaxy, the minihalo is crushed before an appreciable abundance of metals is transported into the primordial gas. If the halo is too far away from the galaxy then there is ample time for metals to move into the cloud, but the shock is too deficient to enrich to the fiducial level.  In most models there is some slight difference in the metal abundance especially for the gas originally found near the center of the dark matter halo. Otherwise, it seems that metal enrichment in these situations is fairly uniform and robust around a value of $\approx 10^{-2}$ Z$_{\odot}$. 

       Thirdly, the observation that globular clusters do not reside within dark matter halos provides a strong constraint on their formation. Such observations show that tidal forces are actively stripping more stars from  globular clusters  (Irwin \& Hatzidimitriou 1993; Grillmair \etal 1995) than would be expected if housed within dark matter halos (Moore 1996; Conroy \etal 2010).  This too is a robust prediction of our model, and in only in a small subset is there any gas left in the dark matter halo.   In fact, only in runs with the largest halo mass, the largest concentration,  and at the largest separations between minihalo and the starburst galaxy was gas retained by the minihalo, and in all these cases at least one other cluster was formed that was unbound from the halo.  Thus outflow-minihalo interactions are a mechanism that primarily, but not exclusively produces dark-matter free clusters. Presumably these rare clusters would still be found with their dark matter halos if they have not been stripped away by some other means or buried within the center of large, low-redshift galaxies.  
              	
	Globular cluster formation has also been studied using simulations of the heirarchical buildup of a Milky Way sized galaxy. Kravtsov \etal (2005) carried out one such a simulation and found that proto-globular clusters are produced in giant molecular clouds within the disk of the galaxy. While this model reproduced many of the properties expected of halo globular clusters, it relied on subsequent violent mergers to move these clusters to the galaxy halo. Muratov \& Gnedin (2010) looked at a similar mechanism that reproduces the observed metallicity distributions found in globular clusters. Griffen \etal (2010) examined a halo from the Aquarius simulation for sites of globular cluster formation, and adopted a simple model in which the cluster sites are determined solely on their temperature. Cluster formation is ended with when the host galaxy is completely reionized. They were able to reproduce the expected number of present day clusters, their positions, and formation ages. However, since their simulation uses only dark matter particles, no mention is made of how the baryonic matter is removed from the dark matter halo or how the metallicity of the resulting cluster arises.   On the other hand, the minihalo-galaxy interactions studied here are both a natural consequence of heirarchical galaxy formation, and they reproduce the masses, metallicities, and dark matter content of halo globular clusters directly in our simulations.
 	
\section{Conclusions}
	
	The early universe was permeated by primordial minihalos that provided the building blocks for larger, later-forming structures. However, since they were not massive enough to form stars on their own, these objects remained passive until acted on by an outside influence.    This means the first galaxies were formed in somewhat larger dark-matter halos, which cooled atomically, formed stars and supernovae, and funned a fraction of the resulting energy  into massive galaxy-sized outflows.    For minihalos in orbit around these early galaxies, such outflows may have triggered a radical transformation.

       In our previous papers, we  studied the role of primordial chemistry and cooling (Paper I) and turbulent mixing and metal-line cooling (Paper II) on outflow-minihalos interactions with a fiducial set of parameters.   In each of these simulations, the baryonic matter was expelled from the minihalo potential and formed dense, cold clusters with nearly constant metal abundances.   Noting that these interactions could be important in understanding the origin of halo globular clusters, we carefully studied each aspect of this interaction and showed that this result was independent of uncertainties in the chemistry and turbulence models and the level of the high-redshift background of dissociating photons.  Furthermore, varying the maximum resolution in our simulations, we showed that medium-resolution ($256 \times 128 \times 128$ effective) simulations were able to faithfully reproduce the outcome of these collisions.

Here we complete this picture and perform a large, medium-resolution parameter study, quantifying the impact of  minihalo mass, minihalo formation redshift, outflow energy, outflow redshift, distance, minihalo concentration, and spin.   For a wide range of parameters, the results are extremely similar. The baryonic matter is expelled from the dark matter halo and formed into at least one dense, cold cluster that is homogeneously enriched with metals.   In fact only under extreme circumstances, such as a large separation between the halo and the galaxy, very low energy outflows, or very high minihalo concentration, is gas retained by the minihalo, and even in these cases at least one other compact, unbound cluster is formed.

Furthermore, our parameter study strengthens the idea that the longest-lived stars formed by this processes will be observable today as members of halo globular clusters.   Like the clusters in our simulations, such globular clusters are observed over a substantial mass range, and their upper mass limit can be directly associated with the maximum minihalo mass, above which atomic cooling becomes efficient.  Over a wide range of energies, redshifts, and distances, outflows are able to accomplish three important jobs necessary to form realistic halo globular clusters: imparting the momentum required to move the pristine gas from the dark matter halo, starting the non-equilibrium chemistry and cooling required for collapse, and providing a source of metals.  The turbulence that follows then mixes these metals into the primordial gas nearly homogeneously.

While the direct detection of outflow-minihalo interactions is beyond current capabilities, it will be well within the reach of the telescopes currently being planned.   Post-processing our simulations we show that  the outflow-driving galaxies are likely to be detectable in broad-band {\em JWST} images, but the clusters themselves are likely to be just beyond their expected detection limits.  On the other hand, narrow-bound imaging of redshifted  Lyman alpha emission from these forming clusters will be well within the capabilities of large ground-based telescopes like the {\em GMT},  {\em TMT}, and {\em E-ELT}.   Such Ly$\alpha$ emitters will appear as bright, extended in a single direction, and pointed directly at larger broad-band detectable starbursts.  This unique signature makes them perfect targets for the next generation of telescopes and an exciting observational probe of an extraordinary mode of high-redshift star formation.

	We are grateful to Aaron Boley, Marcus Br${\rm \ddot{u}}$ggen, and F.X. Timmes for helpful conversations. We acknowledge support from NASA theory grant NNX09AD106. All simulations were conducted on the ÔSaguaroÕ cluster operated by the Fulton School of Engineering at Arizona State University. The results presented here were produced using the FLASH code, a product of the DOE ACS/Alliances funded Center for Astrophysical Thermonuclear Flashes at the University of Chicago.

{}

\end{document}